\documentclass[a4paper,12pt]{article}
\usepackage[authoryear]{natbib}
\usepackage{comment}
\usepackage[utf8]{inputenc}
\usepackage{makeidx}
\usepackage{float}
\usepackage{amssymb}
\usepackage{amsmath}
\usepackage{graphicx}
\usepackage{epstopdf}
\usepackage{authblk}
\usepackage{scalefnt}
\usepackage{bbm}
\usepackage{setspace}
\usepackage[english]{babel}
\usepackage{color}
\usepackage[linkbordercolor={0.5 0.5 0.5}]{hyperref}
\usepackage[open=true, numbered=true]{bookmark}
\usepackage{ragged2e} 
\usepackage{lscape}
\usepackage[section]{placeins} 
\usepackage[top=1.10in, bottom=1.1in, left=1.10in, right=1.10in]{geometry}
\usepackage{setspace}
\usepackage{etoolbox}
\usepackage{booktabs}
\usepackage{adjustbox}
\usepackage[labelformat=simple]{subcaption}
\usepackage{longtable}
\usepackage[normalem]{ulem}
\patchcmd{\thebibliography}{\section*{\refname}}{}{}{}

\usepackage[shrink=60,letterspace=500]{microtype}

\begin{document}

\title{\LARGE Forecasting Macroeconomic Tail Risk in Real Time:\newline Do Textual Data Add Value?\thanks{We thank an associate editor and two referees, the participants of the the 9th Annual Conference of the International Association for Applied Econometrics (IAAE) in Oslo and of the 12th ECB Conference on Forecasting Techniques in Frankfurt, in particular Jasper de Winter (the discussant), for valuable comments. We thank Ignacio Moreira for excellent research assistance. We further gratefully acknowledge the support of the German Research Foundation (project ID 468814087 and project ID 468715873).}}

   \author{Philipp Adämmer\footnote{Institute for Data Science, University of Greifswald E: adaemmerp@uni-greifswald.de} \\
   {\small \emph{University of Greifswald}} \\  \and Jan Pr\"{u}ser\footnote{{Faculty of Statistics, TU Dortmund University, 44221 Dortmund. E: prueser@statistik.tu-dortmund.de}}\\
   {\small \emph{TU Dortmund}} \and Rainer A. Sch\"{u}ssler\footnote{Department of Economics, University of Rostock, Ulmenstra{\ss}e 69, 18057 Rostock.\newline E: raineralexanderschuessler@gmail.com} \\
   	{\small \emph{University of Rostock}}}

\maketitle
\thispagestyle{empty}

\noindent    We examine the incremental value of news-based data relative to the FRED-MD economic indicators for quantile predictions of employment, output, inflation and consumer sentiment in a high-dimensional setting. Our results suggest that news data contain valuable information that is not captured by a large set of economic indicators. We provide empirical evidence that this information can be exploited to improve tail risk predictions. The added value is largest when media coverage and sentiment are combined to compute text-based predictors. Methods that capture quantile-specific non-linearities produce overall superior forecasts relative to methods that feature linear predictive relationships. The results are robust along different modeling choices.

% \sout{Variable importance analyses reveal that tail predictions are determined by both economic and textual indicators, with the latter having the most pronounced impact on consumer sentiment.} \color{blue} 
%\color{red} Wir könnten noch explizit hervorheben, dass Textdaten deutlich weniger Mehrwert für die Mitte der Verteilung haben. Das würde uns von den bisherigen Studien noch deutlicher abgrenzen. \color{black}

 %Finally, We show that the left tails predictions are determined by both textual predictors as well as economic indicators.

\bigskip
\noindent{KEYWORDS:} Quantile Regression, Textual Data, Topic Models, Quantile Regression Forests, Gaussian Processes \\

\bigskip
\noindent{JEL:} C53, C55, E27, E37

\newpage
\doublespacing
%\setstretch{1.52}

\section{Introduction} 
Periods of economic stress such as the Global Financial Crisis or the COVID-19 pandemic have highlighted the importance of tail risk predictions. Quantile predictions of macroeconomic time series provide a more nuanced picture than point forecasts, as the former allow the predictive relationship between the target variable and the covariates to vary across quantiles. Policymakers and central banks are particularly interested in the tails, which are associated with high economic risk. For this reason, the literature on macroeconomic forecasting has paid increasing attention to quantile predictions (see, e.g., \citealp{Manzan2015,korobilis2017,Adrian2019,Carriero2020,Adams2021,clark2022,Pruser2023}).  

%Another recent development in macroeconomic forecasting is the deployment of large data sets, for instance the big data macroeconomic FRED database compiled by  \cite{McCracken2016}, which is used by, e.g., \citealp{kotchoni2019, boot2019,Medeiros2021,Ellingsen2022}). 

Another recent development in macroeconomic forecasting is the use of textual data, which provide timely information that may embed complementary signals to (hard) economic indicators (see e.g., \citealp{larsen2019, Ellingsen2022, Kalamara2022, shapiro2022, vandijk2023, bybee2021}). By extracting information from textual data, researchers seek to quantify the predictive relationship between narratives and economic outcomes. As argued by \cite{Shiller2017}, narratives may even have the power to causally shape economic outcomes. Therefore, quantifying changing narratives based on textual data seems to be a worthwhile endeavour.

Our study takes a data-driven approach to assess whether textual data, relative to a large set of economic indicators, add value to macroeconomic quantile predictions in high-dimensional settings. We focus on monthly tail risk forecasts (nowcasts and one month ahead) of employment, industrial production, inflation and consumer sentiment between 1999:08 and 2021:12. To distinguish between the benefits of text-based predictors for tail predictions compared to the center of the distribution, we look at a broad range of quantiles (5\%, 10\%, 25\%, 50\%, 75\%, 90\%, 95\%).  As data revisions are an important issue in out-of-sample (OOS) macroeconomic forecasting, we use real-time vintages of economic predictors contained in the prominent \cite{McCracken2016} FRED-MD database.

For the computation of text-based predictors, we opt for an unsupervised machine learning approach that is agnostic about the dominant themes in a corpus of documents, namely the Correlated Topic Model (CTM) by \cite{blei2007}. We use the CTM to map nearly 800,000 newspaper articles from \textit{The New York Times} and \textit{The Washington Post} into a set of numerical text-based predictors. The most frequent words within each topic (word cluster) characterize the topic, and the proportion indicates the degree of media coverage of a particular topic at a particular point in time. However, the topic proportions do not indicate whether the tone of the topic is positive or negative. We therefore also use tone-adjusted (sentiment) topic proportions. To investigate whether textual data can be used to improve tail risk forecasts, we use three different sets of predictors: (i) macroeconomic predictors only, (ii) macroeconomic predictors \textit{plus} unadjusted text-based predictors, and (iii) economic predictors \textit{plus} unadjusted text-based predictors \textit{plus} text-based predictors with tone adjustment.\footnote{Text-based predictors are a viable alternative to survey forecasts and other ``soft'' data for quantile predictions. While survey forecasts were found to be valuable in improving point forecasts of macroeconomic time series, they were not found to be helpful in producing more accurate estimates of variance. The reason is that professional forecasters tend to be overconfident, hence providing too narrow density forecasts; see, e.g., \citealp{galvao2021,banbura2021}. Given our focus on quantiles and, in particular, on the tails of the distribution, textual data appear more promising.  For other ``soft'' data such as Google Trends, the available data history is considerably shorter than that of our text-based predictors.}

In terms of forecasting models, two key insights guided our choice. First, mechanisms to prevent overfitting are necessary for OOS forecasting in high-dimensional settings such as ours. We therefore use three Bayesian quantile regressions (QRs) with different shrinkage priors 
as \cite{carriero2022} have shown the importance of using shrinkage for QR in empirical macroeconomics. Second, recent studies point to the importance of capturing  non-linear predictive relationships. For example, \cite{Goulet2022} consider capturing non-linearities as the ``true game changer'' of machine learning methods for macroeconomic forecasting. This is corroborated by  \cite{Medeiros2021} and \cite{clark2022}, who find strong empirical support for tree-based methods. To capture non-linear relationships between the quantile-specific target variables and the covariates, we use non-parametric Bayesian Gaussian Process Regressions \citep{Williams2006} and QR Forests \citep{Meinshausen2006}. %Both methods are equipped with built-in regularization. 
Entertaining both forecasting methods that feature linear and non-linear predictive relationships enables us to evaluate the empirical differences between the model classes. %Following the literature, we evaluate the forecasts with the quantile score or tick loss function \citep{Giacomini2005}.
%assume a linear relationship between the quantile-specific target variables and the covariates
%While most studies analyze the benefits of textual predictors for macroeconomic point forecasts, the role of textual predictors for tail forecasting is largely unexplored. \color{blue} An exception is  \cite{Barbaglia2022}, who use sentiment scores from newspaper articles for predicting key macroeconomic variables and find added value for tail predictions. \cite{Sharpe2023} construct a sentiment index based on the text of the Federal Reserve's Greenbook and assess whether it adds value relative to the point forecasts in the book.\footnote{The Federal Reserve Board staff prepares economic forecasts prior to each FOMC meeting in a document known as the Greenbook (now Tealbook) accompanied by an economic narrative.} \color{blue} Quantile regressions reveal that much of the forecasting power provided by their sentiment measure arises from its signal of downside risks to both economic performance and stock returns. \cite{Filippou2023} also use quantile regressions and find that information in the FOMC narrative about macroeconomic risks and uncertainties is particularly helpful in explaining tail outcomes.\color{black}

While most studies analyze the benefits of text-based predictors for macroeconomic point forecasts, the role of text-based predictors for tail forecasts is largely unexplored. Few exceptions are \cite{Barbaglia2022},  \cite{Sharpe2023} and \cite{Filippou2023}  who use sentiment-based approaches to determine the relevance of textual data (newspaper articles and FOMC speeches) for macroeconomic quantile forecasts. Our approach differs from these studies in three important ways. First, we investigate the added value (if any) of textual data in a setting where the forecaster is working in a data-rich environment using the FRED dataset, which contains a large amount of information about the economy.  We ask whether, in this setting, a large set of text-based predictors can provide additional value for tail risk predictions. Second, we consider not only forecasting models that can deal with high-dimensional predictor sets, but also two methods that can deal with non-linear predictive relationships. Third, we combine unsupervised topic models to extract the content of newspaper articles via estimated topic proportions and further tone-adjust these topic proportions by multiplying them with sentiment scores based on a pre-defined dictionary. By using both unadjusted and tone-adjusted text-based predictors, we can disentangle the added value of textual predictors that embed only information about media coverage from textual data that also includes information about the tone of newspaper articles. Our approach to learning from the news data is agnostic about which themes are relevant for predicting quantiles of macroeconomic times series. This contrasts with approaches such as that of \cite{Barbaglia2022}, who only use text in an article that is semantically dependent on a term of interest (e.g. inflation). 
% Vorschlag von Jan: Apart from the methodological differences, our study also differs in terms of the research question. The studies mentioned above find that text data can lead to improved tail risk forecasts. However, it remains unclear whether these gains would persist if a researcher is already working with a large set of economic indicators. Therefore, a researcher using the FRED data for tail risk forecasting \sout{in a data-rich environment} might ask whether textual data adds further value. This is precisely the research question we \sout{ask} attempt to answer in this paper.

Our key findings can be summarized as follows.  First, textual data contain valuable predictive information that is not embedded in the large set of FRED-MD economic indicators. Our results suggest that text-based predictors can improve tail risk forecasts of macroeconomic variables, but much less for the center of their probability distributions, consistent with the narrative that news signals are most helpful in extreme economic situations. 
%First, textual data contain valuable incremental information particularly useful for
%This finding substantially strengthens previous evidence of the added value of textual data for macroeconomic tail forecasting \citep{Barbaglia2022}, in that we still find incremental predictive power from text-based predictors even relative to a large set of 130 \textit{FRED-MD} economic indicators, and where the textual predictors are selected in a largely data-driven manner rather than having been aggregated before entering the forecasting models.  
 %Second, in our forecasting exercises, adding tone-adjusted text-based predictors leads to substantially lower quantile scores relative to using unadjusted ones only. 
Second, adding tone-adjusted text-based predictors results in more accurate forecast results than using only unadjusted predictors. Third, methods that can capture non-linear predictive relationships produce better predictions than those that cannot.  We have conducted a number of robustness checks that confirm our main findings. Among other things, we find that our results are robust to longer forecast horizons and a range of different choices to produce text-based predictors. \color{black}

% \color{blue} Our results show that textual data contain valuable incremental information that is  particularly useful for predicting the tails of the distribution and less so for the center, consistent with the narrative that timely news signals are most helpful in extreme economic situations. This is generally true for both linear and non-linear models, with the latter providing more accurate quantile forecasts overall. We also find that tone-adjusted textual predictors add further value to the prediction exercises compared to unadjusted ones. \color{black} Our variable importance analyses show that both economic and textual indicators are used for tail predictions. Importantly, combining text and FRED data can yield sizable gains in forecast accuracy and is never much worse than using FRED data only. Moreover, methods which can capture non-linearities produce better now- and forecasts than those which cannot. We finally observe substantial forecast accuracy gains for consumer sentiment when using textual data, which is consistent with the notion that news data are important in forming household expectations \citep{larsen2021}. 

The paper proceeds as follows. Section 2 lays out our methodology. Section 3 outlines our forecasting setup, introduces our predictors, and presents our forecast results. Section 4 concludes. Additional material is relegated to the appendix.\color{black}

\section{Methodology}

\subsection{Forecasting methods}
\subsubsection{Bayesian Quantile Regressions}
In settings with many regressors, Bayesian shrinkage alleviates overfitting
and thus noisy forecasts. Recently, \cite{carriero2022} showed the importance of using shrinkage for QR in empirical macroeconomics.  For Bayesian estimation, we use the mixture
representation established by \cite{yu2001}. For a given variable $y$
of interest that is to be predicted for quantile $\tau $ at horizon $h$, the Bayesian QR can be stated as
\begin{equation}
y_{t+h}={x}_{t}^{\prime}\mathbf{\beta }_{\tau
}+\varepsilon _{\tau ,t+h}\text{,}  \label{bcr0}
\end{equation}

where $\left \{ x_{t}\right \} _{t=1}^{T}$ is a column vector of predictors,  $\beta _{\tau }$ denotes a vector of quantile-specific regression coefficients and $\varepsilon _{\tau ,t+h}$ follows an asymmetric Laplace distribution, see \cite{yu2001}. The dimension of  $x_{t}$ depends on the setting, which is one of the following three possibilities: (i) FRED predictors only, (ii) FRED predictors \textit{plus} unadjusted text-based predictors, and (iii) FRED predictors \textit{plus} unadjusted text-based predictors \textit{plus} text-based predictors with tone adjustment. 

% Using the mixture representation, we can rewrite the QR
% model (\ref{bcr0}) as
% \begin{equation}
% y_{t+h}={x}_{t}^{\prime}\mathbbm{\beta }_{\tau
% }+\theta _{\tau }z_{\tau ,t+h}+\kappa _{\tau }\sqrt{\sigma _{\tau ,h}z_{\tau
% ,t+h}}u_{t+h\text{,}}\label{bcr1}
% \end{equation}

% where $z_{\tau ,t+h}$  is exponentially distributed with scale parameter $%
% \sigma _{\tau ,h};$ $\theta _{\tau }$ and $\kappa _{\tau }$ are
% quantile-specific fixed parameters and $u_{t+h}$ is i.i.d. standard normal; see \cite{yu2001} for further details.

Posterior inference requires specifying a likelihood and eliciting priors
for the coefficients. As Markov Chain Monte Carlo estimation is slow in high dimensions, we use fast variational Bayes approximations for posterior inference. In the interest of brevity we do not give full details
of the estimation, but note that, by introducing auxiliary latent variables, the likelihood is conditionally Gaussian and the errors are conditionally heteroscedastic, see \cite{yu2001} and \cite{Pruser2023}. The shrinkage priors we consider fall into the class of global-local shrinkage priors, where the prior
variance comprises one global term pertaining to all coefficients and
another coefficient-specific term. Our chosen shrinkage priors can be written in the general form 
\begin{equation}
\mathbf{\beta }_{\tau }|\psi _{\tau _{1}},\ldots,\psi _{\tau _{J}},\lambda
_{\tau }\sim \prod \limits_{j=1}^{J}\mathcal{N}\left( 0,\psi _{\tau
j}\lambda _{\tau }\right) ,\text{ }\psi _{\tau j}\sim u,\text{ \ }\lambda
_{\tau }\sim \pi \text{,}  \label{bcr2}
\end{equation}

where $J$ denotes the number of predictors, $\lambda _{\tau }$ denotes a quantile-specific global shrinkage parameter and $\psi _{\tau j}$ are quantile-specific local scaling
parameters that control the coefficient-specific shrinkage intensities.
Different shrinkage priors are generated by choosing different mixing
densities via the functions $u$ and $\pi $. We consider three shrinkage priors that are popular choices in macroeconomic forecasting; see, e.g., \cite{huber2019}, \cite{cross2020}, and \cite{pruser2023DB}:
\begin{itemize}
\item Ridge: The Ridge prior collapses to a purely global shrinkage prior since all local scaling parameters are set equal to $1$. The global shrinkage
parameter follows an inverse Gamma distribution. Formally, we thus have: $
\psi _{\tau j}=1$ \  \ $\forall \tau ,j$ and $\lambda _{\tau }\sim \mathcal{IG
}\left( e_{0},e_{1}\right) $. We choose the hyperparameters $e_{0}=e_{1}=0$ which results in a flat prior, see \cite{tipping2001sparse}. Overall, the Ridge prior offers a comparatively low degree of flexibility for variable-specific deviations from the global shrinkage pattern. This prior is supposed to work well when many predictors are relevant, being consistent with a dense representation of the prediction problem.
\item Horseshoe: In contrast to the Ridge prior, the Horseshoe prior \citep{Carvalho2010} offers variable-specific shrinkage. The Horseshoe sets both $u$ and $\pi$ to half-Cauchy distributions: $\sqrt{\psi _{\tau j}}\sim \mathcal{C}^{+}\left( 0,1\right) $ and 
$\sqrt{\lambda _{\tau }}\sim \mathcal{C}^{+}\left( 0,1\right)$. An advantage of the Horseshoe prior is that it does not require the researcher to elicit any tuning parameters. The Horseshoe prior has been shown to have excellent posterior contraction properties, see, e.g., \cite{ghosh2016}. The Horseshoe prior spikes at zero and has fat tails. Hence, it is supposed to shrink small coefficients of unimportant predictors to zero, but (in relative terms) large coefficients of the informative predictors are not shrunk much.  Accordingly, the Horseshoe prior should work well when only a small number from the pool of predictors is useful, consistent with a sparse representation of the prediction problem. 

\item Lasso: The Lasso prior is a special case of the Normal-Gamma prior of \cite{brown2010}. The Lasso 
involves setting $u$ and $\pi $ to Gamma
distributions: $\psi _{\tau j}\sim \mathcal{G}\left( 1,\lambda _{\tau
}/2\right) $ and $\lambda _{\tau }\sim \mathcal{G}\left( c_{0},d_{0}\right) $. We set the hyperparameters $c_{0}=d_{0}=0$.
%Similar to Horseshoe, the Lasso prior can shrink coefficients to zero, thereby performing variable selection and thus encouraging sparsity.  % implying heavy global shrinkage. The marginal prior of the coefficients exhibits fat tails. %Overall, the Lasso prior offers richer shrinkage patterns than Horseshoe and Ridge.
\end{itemize}

\subsubsection{Gaussian Process Regressions}
Gaussian Process Regression \citep{Williams2006} is a non-parametric method that was recently used for inflation forecasting by \cite{clark2022f}. It elicits a process prior on the function $g_{\tau }\left({x}_{t}\right):$
\begin{equation}
g_{\tau }\left({x}_{t}\right) \sim \mathcal{GP}\left( \mu _{\tau
}\left( {x}_{t}\right) ,\mathcal{K}\left( {x}_{t},{x}_{
\mathfrak{t}}\right) \right) \text{,}\label{GP0}
\end{equation}

where we set the mean function $\mu _{\tau }\left( {x}_{t}\right) $
to zero. The kernel function $\mathcal{K}\left( {x}_{t},{x}_{
\mathfrak{t}}^{^{\prime }}\right) $ describes the relationship between $
{x}_{t}$ and ${x}_{\mathfrak{t}}$, for $t$, $\mathfrak{t=}
1,\ldots,T$.

As ${x}_{t}$ is observed at discrete points in time, ${g}
_{\tau }=\left( g_{\tau }\left( {x}_{1}\right),\ldots,g_{\tau }\left( 
{x}_{T}\right) \right) ^{\prime }$:
\begin{equation}
{g}_{\tau }\sim \mathcal{N}\left( {0}_{T},{K}\left( 
{w}\right) \right) \text{, }  \label{gprior}
\end{equation}

where ${K}\left( {w}\right) $ refers to a $T\times T$-dimensional matrix with $\left( t,\mathfrak{t}\right) $-th element $
\mathcal{K}\left( {x}_{t},{x}_{\mathfrak{t}}\right) $.

The type of kernel determines the estimated function. We choose a squared
exponential kernel
\begin{equation}
\mathcal{K}\left( {x}_{t},{x}_{\mathfrak{t}}\right)
=w_{1}\times e^{-\frac{w_{2}}{2}\left \Vert {x}_{t}-{x}_{
\mathfrak{t}}\right \Vert ^{2}}\text{,}  \label{kernel0}
\end{equation}

where we follow \cite{chaudhuri2017} for setting the hyperparameters ${w=}
\left( w_{1},w_{2}\right) ^{^{\prime }}$, which govern the smoothness of the function.

Above we have outlined  the function-space view of the GP regression. In the
alternative weight-space view, which is convenient for estimation, the GP regression can be expressed as 
\begin{equation}
{y}={Z\gamma }_{\tau }+{\varepsilon }\text{,}\label{wspace}
\end{equation}

where ${y}$ denotes the stacked dependent variables, ${Z}$
represents the lower Cholesky factor of ${K}$, and $\mathbf{
\gamma }_{\tau }\sim \mathcal{N}\left( {0}_{T},{I}_{T}\right) $.

\subsubsection{Quantile Regression Forests}

Our last forecasting method is a frequentist non-parametric method, namely QR Forests, an extension of Random Forests \citep{breiman2001} for conditional point estimation to conditional quantile estimation based on an ensemble of trees \citep{Meinshausen2006}. Random Forests and QR Forests capture non-linear predictive relationships and, especially due to this feature, have been found to perform well in macroeconomic forecasting; see, e.g., \citealp{Medeiros2021,clark2022}). 

Random Forests grow a large number of trees by using $t$ observations:
\begin{equation*}
\left( Y_{i},X_{i}\right) ,i=1,\ldots,t\text{,}
\end{equation*}

where $Y$ is the variable of interest and $X$ is a (possibly
high-dimensional) predictor variable. For ease of notation we drop time subscripts. For each tree and node, Random
Forests uses a random subset of predictors to split on.\footnote{In our empirical work, for a $p$-dimensional predictive variable, at each node we use the default choice of $\sqrt{p}$ randomly selected predictors.} The intuition of this random selection is to de-correlate the trees and thus to decrease the variance of the forecasts. In Random Forests, the conditional mean prediction of $Y$, given $X=x$, is generated as the weighted sum over all observations:
\begin{equation}
\widehat{\mu }\left( x\right) =\sum \limits_{i=1}^{t}w_{i}\left( x\right)
Y_{i}\text{,}  \label{cmean}
\end{equation}

where the weights $w_{i}\left( x\right) $ are computed over the collection
of trees. In each tree, the conditional mean prediction is the simple
average of all observations that fall into the same leaf when dropping down $x$; the remaining observations are neglected.

\cite{Meinshausen2006} extended the Random Forests to QR Forests. The conditional distribution function of $Y$, given $X=x$, is
\begin{equation}
F\left( y|X=x\right) =P\left( Y\leq y|X=x\right) =\mathbbm{E}\left( \mathbbm{1}_{\left \{ Y\leq
y\right \} }|X=x\right) \text{.}  \label{conditional}
\end{equation}
Instead of approximating the conditional mean $\mathbbm{E}\left( Y|X=x\right)$ in case of Random Forests, for QR Forests, $\mathbbm{E}\left( \mathbbm{1}_{\left \{ Y\leq y\right \} }|X=x\right) $
is approximated by the weighted mean over the observations $\mathbbm{1}_{\left \{ Y_{i}\leq y\right \}}$,
\begin{equation}
\widehat{F}\left( y|X=x\right) =\sum \limits_{i=1}^{t}w_{i}\left( x\right)
\mathbbm{1}_{\left \{ Y_{i}\leq y\right \} }\text{,}  \label{dist}
\end{equation}

where $w_{i}\left( x\right) $ are the same weights as for Random
Forests. Based on $\widehat{F}\left( y|X=x\right) $, we can estimate the
desired conditional $\alpha $-quantile $Q_{\alpha }\left( x\right) $ as

\begin{equation}
\widehat{Q}_{\alpha }\left( x\right) =\inf \left( y:\widehat{F}\left(
y|X=x\right) \geq \alpha \right) \text{.}  \label{quantile}
\end{equation}

An important difference between QR Forests and Random Forests is that, for each node in each tree, Random Forests keep only the mean of the observations that fall in that node, whereas QR Forests keep the values of all the observations in that node to compute the conditional distribution.

\subsection{Probabilistic topic models}
We use a probabilistic topic model to create text-based predictors. Topic models are data-driven and use a probabilistic approach to identify themes within a large set of written documents. Unlike dictionary-based and Boolean approaches, topic models are initially context-agnostic, finding word clusters (topics) solely based on the co-occurrence of words.

%The most prominent topic models are unsupervised, meaning that no additional information beyond the texts themselves are needed. In addition, documents are treated as ``bag of words'', which enables to represent a document as a vector of word counts.

%The values within a document-term matrix (dtm) count how often a certain word (column) occurs within a certain document (row). %Even though word order and grammar may be important to fully understand the meaning of a text, they are not important for identifying general themes within a collection of documents.

The most prominent topic model is latent Dirichlet allocation (LDA) by \cite{blei2003}. It posits that documents are generated by a stochastic process where each text is a mixture of latent topics and each topic is a probability distribution over the same vocabulary, but with different probabilities for each word. %\color{red} For example, a topic about ``inflation'' may assign high probabilities to words such as ``commodity'' and ``price'', whereas a topic about ``financial markets'' may assign high probabilities to words like ``stock'' and ``yield``. \color{black}
%Even though all documents share the same set of topics, each document has its unique mixture of them. 
Despite its prominence and advantages, Latent Dirichlet Allocation (LDA) cannot account for the fact that certain topics tend to co-occur together within documents (e.g., inflation and commodity prices). We therefore use the more sophisticated CTM by \cite{blei2007}, which was developed to address this limitation.\footnote{Economic studies have predominantly relied on LDA by \cite{blei2003} to extract news-based predictors. These predictors have then been used, for instance, to investigate the value of news data for modeling macroeconomic dynamics \citep{larsen2019, bybee2021}, to construct a daily business cycle index \citep{thorsrud2020}, to predict US macroeconomic variables \citep{Ellingsen2022}, and to nowcast US GDP \cite{Babii2021} as well as Chinese GDP and inflation expectations \citep{zheng2023}. The CTM and further extensions have been used to produce topic proportions which have then been used, for example, to analyze news coverage of China \citep{roberts2016},  to investigate the impact of presidential tax speeches on economic activity \citep{dybowski2018}, to forecast the equity premium \citep{adaemmer2020}, and to analyze European Central Bank communication \citep{dybowski2020, bohl2023}. \color{blue} \color{black}} %\color{red} For example, an article about inflation is more likely to also touch on other topics related to macroeconomics and finance than on topics related to, say, entertainment.\color{black}

\begin{figure}[H]
\begin{center}
\includegraphics[width = \linewidth]{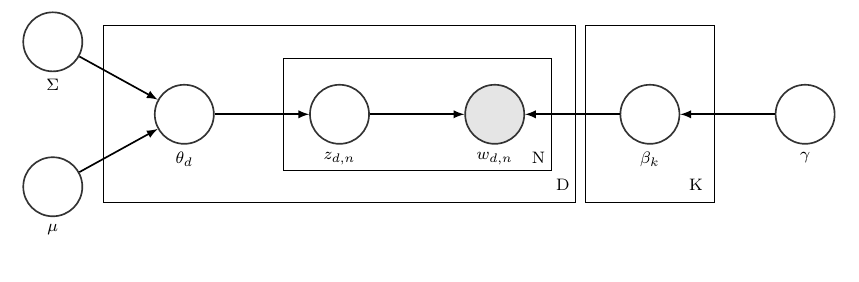}
\caption{\footnotesize Generative process of the CTM based on \cite{blei2007}.}
\label{GM_CTM}
\end{center}
\end{figure}

Figure \ref{GM_CTM} shows a graphical model of the generative process assumed by the CTM, where edges denote dependencies, nodes are random variables, and plates are replicated variables. The only observable variables of the model are the words ($w$).

The key difference between LDA and the CTM is the assumption for the topic proportions, $\theta_d$. The CTM assumes that topic proportions follow a logistic normal distribution---which has a mean vector ($\mu$) and a covariance matrix ($\Sigma$)---as opposed to LDA, which assumes that topic proportions come from a Dirichlet distribution. Consistent with the more realistic assumptions of correlating topics, the CTM yields higher values for statistics that measure the predictive performance regarding unseen documents \citep{blei2007}.
The generative process of the CTM can be written as follows (local mathematical notation):

\begin{enumerate}
    \item The probabilities of words in topic \emph{k} are given in vector $\beta_{k}$. Each topic \emph{k} is a multinomial distribution over a unique vocabulary, generated from the Dirichlet distribution with input vector $\gamma$:  \vspace{-.25cm}
        $$\beta_k \sim Dir(\gamma).$$ 
    \item For each document \emph{d} and topic \emph{k}, topic proportions are drawn from the logistic normal distribution: 
        \begin{equation*} 
        \begin{aligned}
        \eta_d &\sim \mathcal{N}(\mu, \Sigma) \\
        \theta_{d, k} & =  \frac{\exp{(\eta_{d, k})}}{\sum_{i=1}^K \exp{(\eta_{d, i})}},
        \end{aligned}
         \end{equation*}
         where \emph{K} denotes the number of topics.  For each document \emph{d}, we obtain the \emph{K}-dimensional vector of topic proportions as $\theta_{d}=(\theta_{d,1},\ldots,\theta_{d,K})'$.
         \begin{enumerate}
            \item For each word $n$, a topic assignment is drawn from the multinomial distribution:
            $$z_{d, n} | \theta_d \sim Mult(\theta_d). $$
             \item Each word is drawn from the multinomial distribution:
             $$w_{d, n} | z_{d, n}, \beta_{1:K} \sim Mult(\beta_{z_{d, n}}). $$         
        \end{enumerate}
\end{enumerate}

We use the \emph{partially collapsed variational EM algorithm} by \cite{roberts2016} to estimate the CTM. The approach is implemented in the \texttt{R}-package \texttt{stm} by \cite{roberts2019}. We are particularly interested in $\theta_d$, namely the topic proportions of each document, whose aggregates serve as our text-based predictors; see Section \ref{text}.

%The plates are replicated variables, namely topics ($K$), documents ($D$) and the words within each document ($N$). The topics $\beta_{1:K}$ are shared by all documents and they are assumed to be drawn from a Dirichlet distribution with prior $\gamma$. The topic proportions $\theta_d$ are drawn for each document $d$ from a logistic normal distribution with mean vector $\mu$ and covariance matrix $\Sigma$. $z_{d,n}$ is the so-called \emph{topic assignment}, which is drawn from a multinomial distribution with probability vector $\theta_d$. Each document's word $w_{d,n}$ is thus drawn from topic $z_{d,n}$, whose probabilities are given in $\beta_{z_{d,n}}$. 
%The goal of the algorithm is to uncover the hidden structure that has most likely generated the observed words within each document. Consequently, it estimates the topic distributions $\beta_{1:K}$ and the topic mixtures $\theta_d$. To estimate the CTM, we use the efficient \emph{partially collapsed variational EM algorithm} by \cite{roberts2016}, which is implemented in \cite{roberts2019}. Apart from its speed, the approach also uses a deterministic ``spectral initialization'' method, making the results reproducible.

\section{Empirical Work}

\subsection{Forecasting setup}\label{setup}

We generate monthly quantile forecasts (nowcasts and one month ahead) for employment, inflation (Total CPI), industrial production, and consumer sentiment. We choose employment, inflation and industrial production as our target variables because they are key macroeconomic variables used in several forecasting studies; see, e.g. \citealp{Kalamara2022, Barbaglia2022}. Consumer sentiment is used to assess whether news data are important in forming household expectations \citep{larsen2021}.  Depending on the setting, we incorporate FRED predictors, text-based predictors, or both together. We detail our predictors in Section \ref{predsets}. In addition, all model specifications include 12 lags of the respective (transformed) variable of interest.

For our target variables and the FRED predictors, we use vintage data from the \cite{McCracken2016} database. We use 98 indicators from the FRED-MD database and transform the data to achieve stationarity \citep{McCracken2016}.\footnote{The transformation of the variables can be found in Appendix \ref{AppC}. We use all indicators from the FRED-MD database that are available both at the beginning and at the end of our out-of-sample period.} Our estimation sample starts in 1980:06, when the text-based predictors start. We run recursive estimations based on an expanding window. Our evaluation period runs from 1999:08 to 2021:12.

%Therefore, we use all news articles from a given month to calculate the topic proportions used for predictions. \color{black} For nowcasts of a given month, w

We compute our predictions at the end of a given month. Due to publication lags, we include macroeconomic predictors from the previous month, while we include text-based and financial predictors from said month.\footnote{See Appendix \ref{AppC} for which variables are classified as financial predictors. Note that few indicators are lagged by more than one month, and we adjust for this.} For example, if we are at the end of December and produce a nowcast for December, we use the macroeconomic predictors from November released in December, and the financial and text-based predictors from December. Similarly, for one-month-ahead forecasts, if we are at the end of December and produce a prediction for January, we use the macroeconomic predictors from November released in December and the financial and text-based predictors from December.

Our five forecasting models are three versions of Bayesian QRs with different shrinkage priors (Horseshoe, Ridge and Lasso), QR Forests and Gaussian Process Regressions. \color{black}

We evaluate our forecasting models with the quantile score (QS), which %or tick loss function \citep{Giacomini2005}. The quantile score (QS) 
is computed as
\begin{equation}
QS_{\tau ,t+h}=\left( y_{t+h}-Q_{\tau ,t+h}\right) \left(
\tau -\mathbbm{1}_{\left \{ y_{t+h}\leq Q_{\tau ,t+h}\right \}
}\right) \text{,}  \label{qscore}
\end{equation}

where  $y_{t+h}$ is the actual outcome of the variable of interest in $t+h$, $Q_{\tau ,t+h}$ denotes the forecast of quantile $\tau$ for $t+h$. The indicator function $\mathbbm{1}_{\left \{ y_{t+h}\leq Q_{\tau ,t+h}\right \}}$ takes on a value of $1$ if the outcome is not higher than the quantile forecast, and $0$ otherwise.

\subsection{Predictor sets}\label{predsets}

\subsubsection{Economic predictors} 
%Our macroeconomic predictors are from the \cite{McCracken2016} \textit{FRED-MD} database. 
We select all macroeconomic predictors from the \cite{McCracken2016} FRED-MD database that are available at both the end and the beginning of the sample. The predictors can be classified into eight categories: (i) output and income, (ii) labor market, (iii) housing, (iv) consumption, orders, and inventories, (v) money and credit, (vi) interest and exchange rates, (vii) prices, and (viii) stock market. For the classification of the variables, see \url{https://www.ssc.wisc.edu/~bhansen/econometrics/FRED-MD_description.pdf}.

\subsubsection{Text-based predictors}\label{text}
We used the legal database LexisNexis to download 793,013 economically related newspaper articles from \textit{The New York Times} and \textit{The Washington Post} between 1980:06 and 2021:12. %\footnote{For instance, we looked for the string \emph{econom} in the texts' bodies and headlines.} 
We then conducted Part-of-Speech-Tagging \citep{benoit2020} to remove anything but nouns from the articles. Our reason for this choice is that topic models aim to summarize the \emph{content} of documents, which is predominantly described by nouns, in contrast to sentiment analysis which aims to describe the documents' \emph{tone}. In addition, \cite{martin2015} found highest values of semantic coherence when using a nouns-only approach, a metric that strongly correlates with human judgement.

We tokenized the documents into single words, removed punctuation, numbers, symbols, stopwords, etc., and constructed a document-term matrix (dtm). A dtm counts how often a certain word (column) occurs within a certain document (row). We then computed term-frequency inverse-document-frequency (tf-idf) values for each word to extract the 10,000 most relevant terms until 1999:08. The final dtm served as the input for the CTM. In addition, the number of topics $K$ has to be given as an input by the researcher. \cite{larsen2019, thorsrud2020, Ellingsen2022} use LDA to compute topic proportions for different inference and prediction tasks. In each study, they choose 80 topics. We follow them and use $K=80$ as our baseline model.\footnote{In Appendix \ref{topalt}, we provide robustness checks for the number of topics, which are based on topic interpretability \citep{roberts2019,adaemmer2020} and an automated search approach by \cite{Mimno2014}. Overall, our main findings are robust to alternative choices of the number of topics.} 

%The only input for the CTM that has to be given by the researcher is the number of topics $K$. 

We estimated for each document on each day 80 topic proportions based on the newspaper articles until 1999:08. After then, we computed the topic proportions OOS to exclude any look-ahead bias, similar to \cite{Ellingsen2022}. Finally, for a given month, we computed the simple average over all documents' estimated topic proportions. Note that $\theta_{d,k}$ denotes the (estimated) topic proportion of the $k$-th topic for the $d$-th document. For a given month $m$ and for each topic $k$, we compute the average topic proportion as follows: %of topic proportions over all documents that fall into the respective month: 
\begin{equation}
\theta_{k}^{(m)} = \frac{1}{D_m}\sum \limits_{d=1}^{D_m}\theta_{d,k},
\label{formula_tm}
\end{equation}
where $D_{m}$ denotes the number of documents that fall into the $m$-th month. Applying (\ref{formula_tm}) to all \emph{K} (estimated) topic proportions produces a $K\times1$ vector of (estimated) topic proportions for the $m$-th month, that is $\theta^{(m)}=(\theta_{1}^{(m)},\ldots,\theta_{K}^{(m)})'$.

 Topic proportions only reveal information about the content of the newspaper articles, not about the tone that accompanies them. However, the tone may contain additional predictive information. We therefore also consider tone-adjusted (sentiment) topic proportions as predictors. To do so, we rely on the dictionary compiled by \cite{Barbaglia2022}, where each word contains a sentiment value between -1 and 1. The more positive (negative) the sentiment score, the more positive (negative) is the tone of the respective word. Note that we use all the words in the sentiment dictionary to calculate a document's sentiment score, unlike the nouns-only approach for calculating the topic proportions. The sentiment score (\textit{Sent}) for the $d$-th document is computed as follows:
\begin{equation}
Sent_{d} = \frac{\sum\limits_{i=1}^N Score_i}{\#\text{Sentiment Words in \emph{d}}},
\label{sent0}
\end{equation}

where $Score_i$ is the sentiment score for the $i$-th word in the respective document. The number of words in each article is denoted by $N$. The sum of a document's sentiment score is normalized by the number of words with an associated sentiment score in the respective text. Figure \ref{Sent_figure} shows  our monthly sentiment time series, computed as $Sent^{(m)} = \frac{1}{D_m}\sum \limits_{d=1}^{D_m}Sent_{d}$. Gray-shaded areas indicate NBER-dated recessions. The time series captures well economic periods of booms and busts. 

\begin{figure}[H]
\begin{center}
\includegraphics[width = \linewidth]{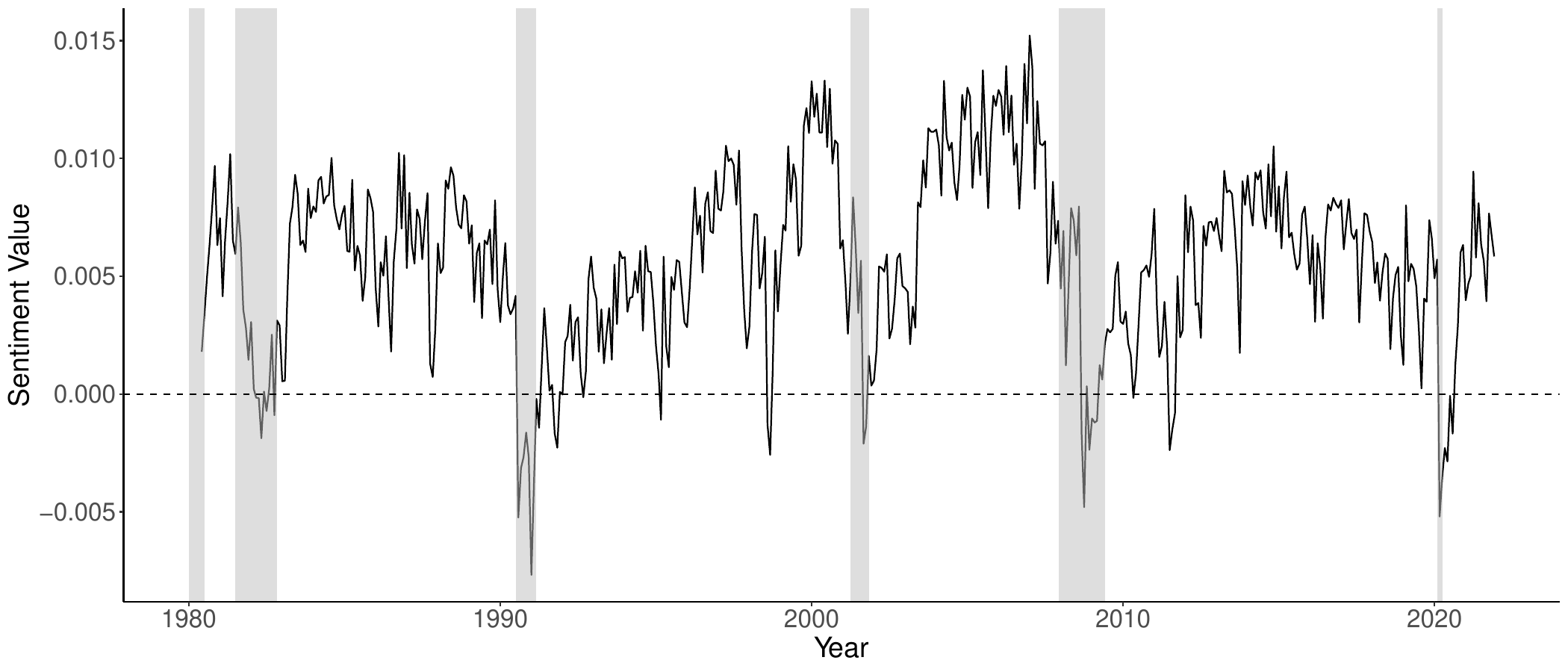}
\caption{\footnotesize Monthly sentiment scores based on the dictionary compiled by \cite{Barbaglia2022}. Gray-shaded areas indicate NBER-dated recessions.}
\label{Sent_figure}
\end{center}
\end{figure}

We use the sentiment score of each document to adjust each topic proportion:\footnote{\cite{Ellingsen2022} also tone-adjust their topic proportions, but in a different manner. In addition, they use the \emph{Harvard IV-4 Psychological Dictionary}, which only distinguishes between positive and negative words and was not designed for economic and financial applications.}
\begin{equation}
\text{SentTopic}_{d,k} = \theta_{d,k}\times Sent_{d}.
\label{sent}
\end{equation}

We then compute the average over all documents of a given month \emph{m}: 
\begin{equation}
\text{SentTopic}_{k}^{(m)} = \frac{1}{D_m}\sum \limits_{d=1}^{D_m}\text{SentTopic}_{d,k}.
\label{sentavg}
\end{equation}

The first row of Figure \ref{Sent_figure} shows the trajectories of three selected topics without any tone adjustment. A high value for a certain topic at a given point in time indicates high media coverage of said topic at that point in time.\footnote{ The labels are based on the top five words of each topic. We emphasize that topic labeling is notoriously subjective, which is why we show the top five words of all topics in Appendix \ref{AppA}.} The figures show that our topics capture important political and economic events such as several debt crises in Mexico, Asia and Europe (Topic 9), the beginning of the Gulf War in 1991 (Topic 27) and the Great Recession which emanated in the housing market (Topic 46).   

\begin{figure}[H]
\begin{center}
\includegraphics[width = \linewidth]{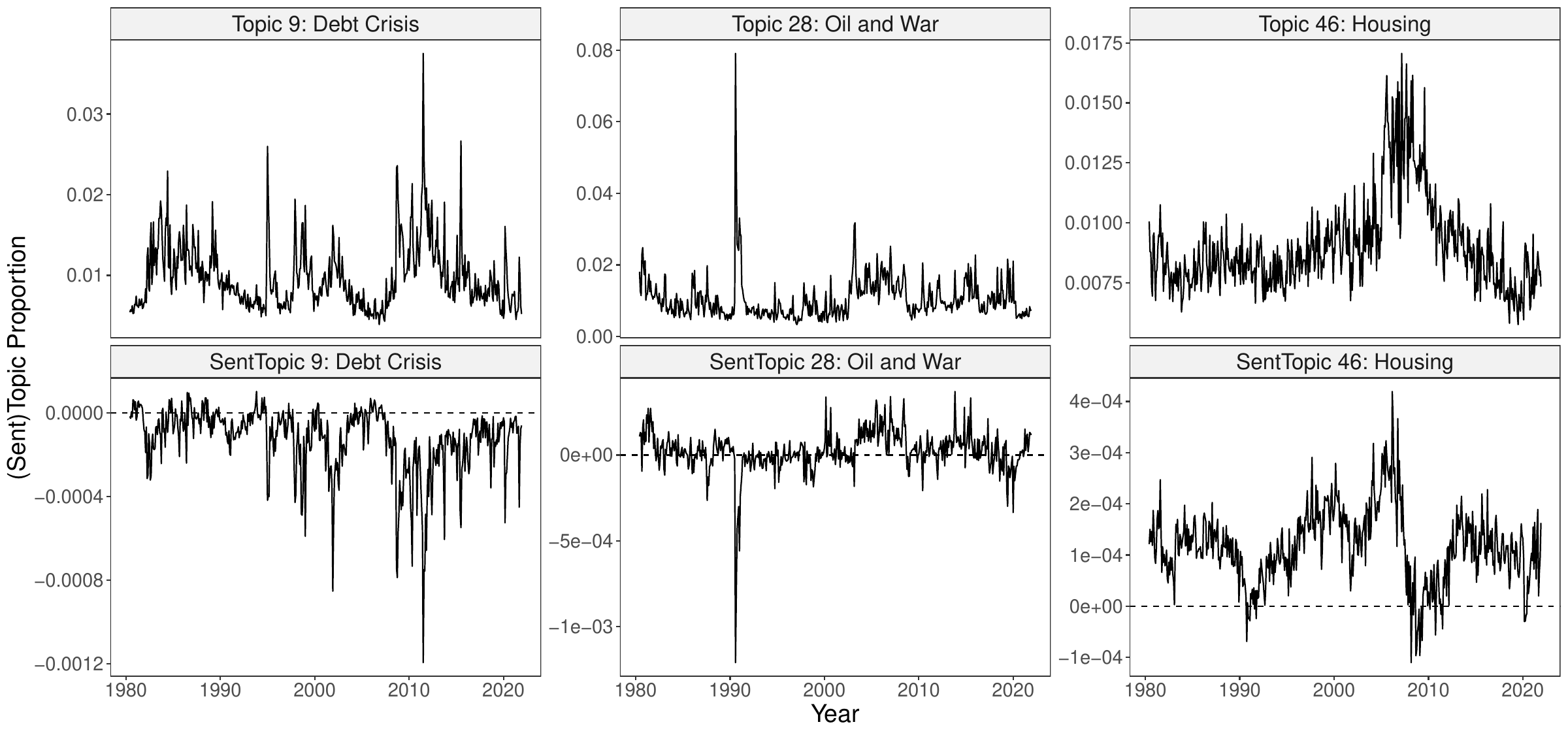}
\caption{\footnotesize Selected monthly unadjusted topic proportions (first row) and with tone-adjustment (second row) based on the dictionary of \cite{Barbaglia2022}.\color{black}}
\label{Sent_figure}
\end{center}
\end{figure}

The second row of Figure \ref{Sent_figure} shows the tone-adjusted topic proportions, which reflect the severity of several events. For example, the tone in articles about debt crises (9) was overall negative. The series of SentTopic 28 drops sharply at the beginning of the Gulf War in 1991 and SentTopic 46, which mainly covers issues related to the housing market, clearly changes in sentiment from positive to negative as the collapse of the US housing market became apparent during the Great Financial Crisis. We use both the monthly averages of unadjusted topic proportions (\ref{formula_tm}) and with tone adjustment (\ref{sentavg}) as text-based predictors in our empirical analyses.

%\subsubsection{Robustness checks with textual predictors}\label{robustness_text}

%\sout{We have also taken advantage of the fact that textual data are available at a higher frequency than macroeconomic data. More specifically, we computed separate topic proportions for each week of the month and for each topic. In the case of $K=80$ this gave us $4\times80=320$ predictors. Interestingly, we did not find any added value in doing so (unreported results), suggesting that it is very difficult to identify when the news are relevant during the month. It might be interesting to see in future research whether tone-adjusted topic proportions add value for predicting the tails.}\color{black}

\subsection{Forecast results}
We first focus on the results for the linear Bayesian QR models with different shrinkage priors, namely Horseshoe, Lasso and Ridge. Our primary interest here is to assess the added value (if any) of the text-based predictors. We then assess how methods that use the same sets of predictors but allow for non-linear predictive relationships (QR Forests and Gaussian processes) perform against the overall most successful linear model.

\begin{figure}[H]
\centering
\begin{subfigure}[b]{.78\textwidth}
    \subcaption[short for lof]{(Nowcast, h = 0)}
   \includegraphics[width=1\linewidth]{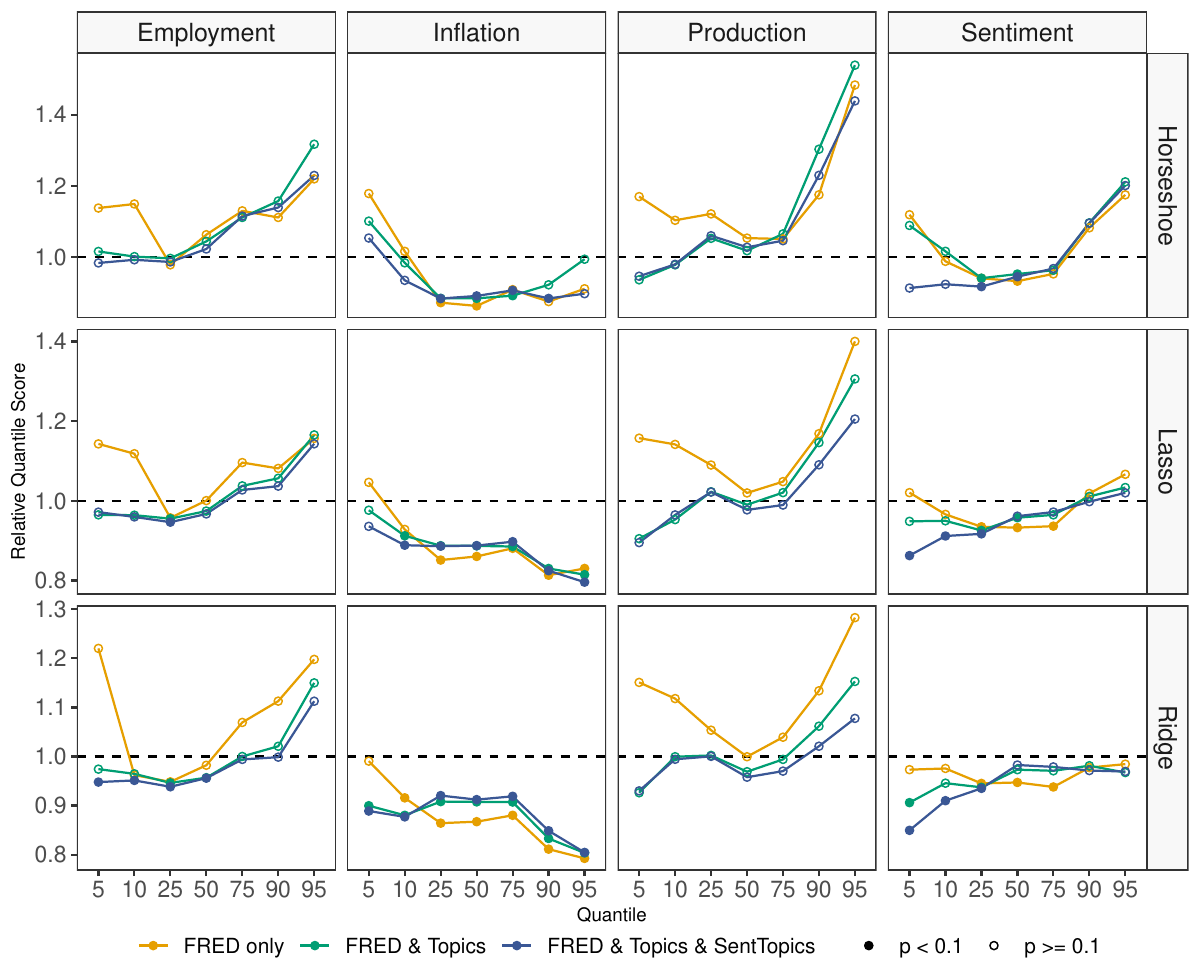}   
   \label{fig:Ng1} 
\end{subfigure}

\begin{subfigure}[b]{.78\textwidth}
\subcaption[short for lof]{(Forecast, h = 1)}
   \includegraphics[width=1\linewidth]{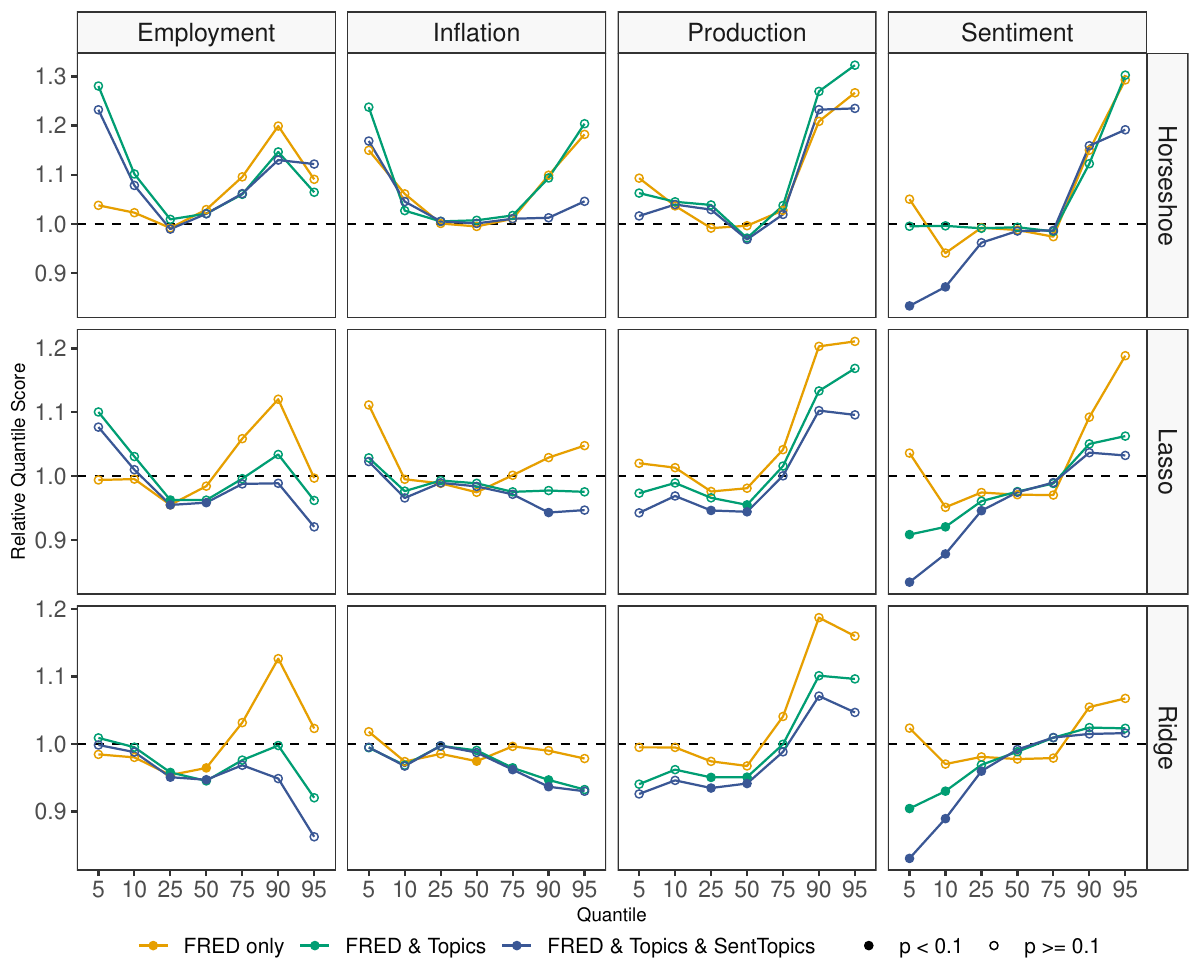}
   \label{fig:Ng2}
\end{subfigure}
\vspace{-.5cm}
\caption{\footnotesize Evaluation of quantile nowcasts (h=0) and quantile one month ahead forecasts (h=1) for the Bayesian QRs with different shrinkage priors (Horseshoe, Ridge and Lasso). The dotted black horizontal line shows the quantile score of the quantile AR(1) benchmark which is standardized to 1.0. Scores below (above) 1.0 indicate more (less) precise forecasts for a given quantile compared to the quantile AR(1) benchmark. Inside colored dots indicate significantly higher forecast accuracy compared to the quantile AR(1) benchmark according to a one-tailed \cite{diebold1995} test at the $10\%$ level.}
% \caption[]{\scriptsize Linear models. One-step-ahead (h = 1) quantile scores.}
\label{linear}
\end{figure}

For Horseshoe, Lasso and Ridge, Figure \ref{linear} summarizes the results for nowcasts and one month ahead forecasts, respectively. Each column denotes a target variable and each row a (linear) model. Forecast accuracy is assessed by quantile scores across different quantiles $\tau$: $\tau$ = 5\%, 10\%, 25\%, 50\%, 75\%, 90\%, 95\%. A quantile AR(1) model serves as our benchmark.\footnote{More precisely, we compute a (frequentist) quantile AR(1) model. For example, in case of a nowcast at the end of December (for December), the target variable's latest available value from November is used as the single predictor in the quantile AR(1): $y_{December}=\beta_{0,\tau}+\beta_{1,\tau}y_{November}+\varepsilon_{December}$.} Relative quantile scores below (above) one indicate more (less) precise forecasts compared to the benchmark model. 
%\color{red} We have further extended AR(1) with dynamically extracted factors, but found the AR(1) model to be the most competitive benchmark.\color{black}

%Quantile scores below (above) one indicate more (less) precise forecasts compared to a linear quantile regression with a constant term and one lag of the variable of intrest (AR(1) bechmark) for a given quantile. 

The combination of macroeconomic predictors and text-based predictors leads to accuracy gains in many cases, especially in the tails, compared to the setting with macroeconomic predictors alone. The addition of tone-adjusted topic proportions leads to even lower overall quantile scores compared to economic predictors and unadjusted topic proportions (FRED \& Topics). In those cases where the addition of text-based predictors does not improve forecast performance, it does not have a materially adverse effect either. The incremental value from adding text-based predictors is pronounced in the tails. Improved tail forecasting with textual data suggests a scenario where text-based predictors are particularly helpful in extreme economic environments, where timely news data can reflect the overall economic picture and be useful predictors of macroeconomic variables. In times such as the Great Recession or the COVID-19 pandemic, uncertainty and political change may have been better captured than by hard economic indicators. 

Within the class of the three linear models, the Horseshoe shrinkage prior underperforms the Lasso and Ridge in the tails. Ridge, as a purely global shrinkage prior, outperforms Horseshoe and Lasso. Thus, the richer shrinkage patterns offered by Horseshoe does not pay off in our analysis. These results are consistent with a dense representation of the prediction problem, where many weak predictors are relevant, rather than a sparse structure, where a few important predictors can be selected and the others are discarded.  

Ridge produces more precise nowcasts for the left tails of employment and production and for both tails of inflation and consumer sentiment relative to a quantile AR(1) model when using FRED \& Topics \& SentTopics. Ridge is also successful in one month ahead forecasting left-tail outcomes of production and the right-tail outcomes of inflation and employment when using FRED \& Topics \& SentTopics. %Hence, (tone-adjusted) textual predictors are found to be most helpful for downside economic risks when using Ridge in this forecasting exercise. 
For consumer sentiment (nowcasts and one month ahead forecasts), quantile scores markedly improve in the left tail when (in particular tone-adjusted) text-based predictors are added. This result is interesting from an information processing aspect, supporting the notion that news data play an important role for households in forming their expectations \citep{larsen2021}.

\begin{figure}[H]
\centering
\begin{subfigure}[b]{.83\textwidth}
    \subcaption[short for lof]{(Nowcast, h = 0)}
   \includegraphics[width=1\linewidth]{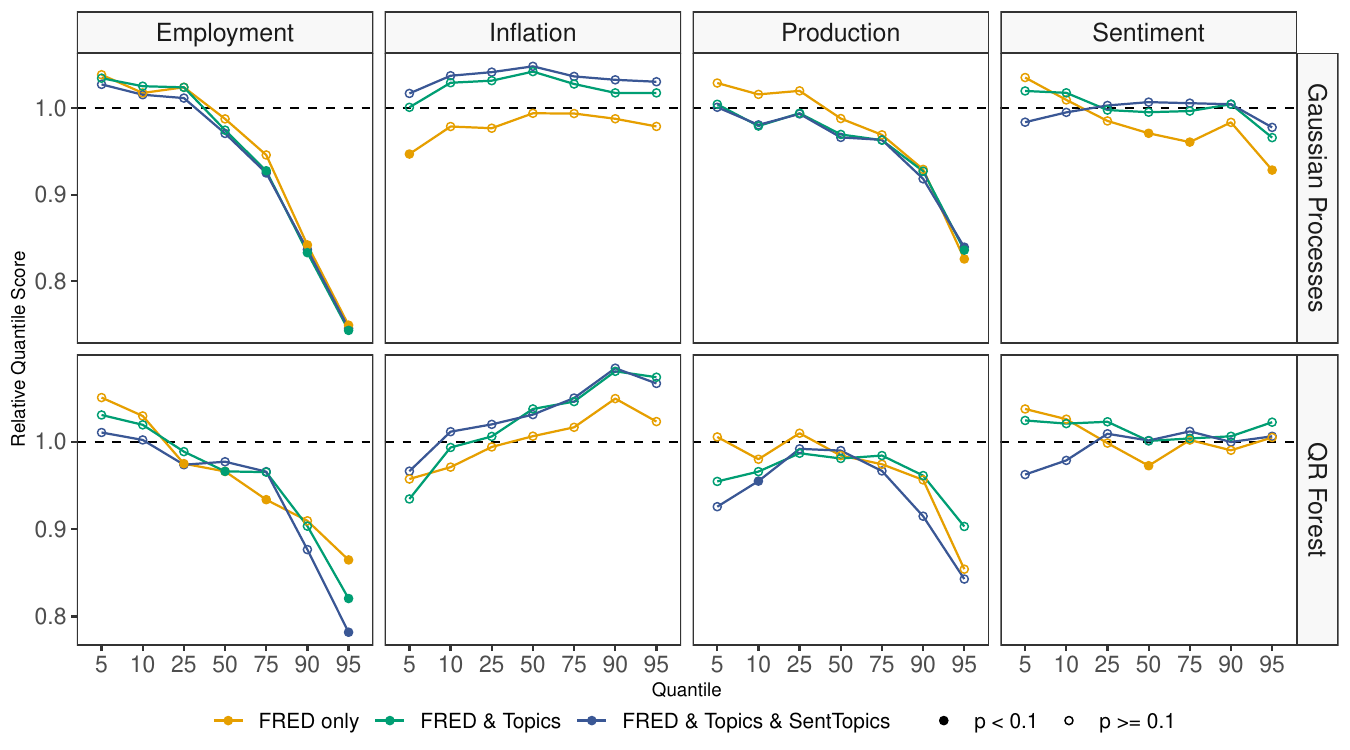}   
   \label{fig:Ng1} 
\end{subfigure}

\begin{subfigure}[b]{.83\textwidth}
\subcaption[short for lof]{(Forecast, h = 1)}
   \includegraphics[width=1\linewidth]{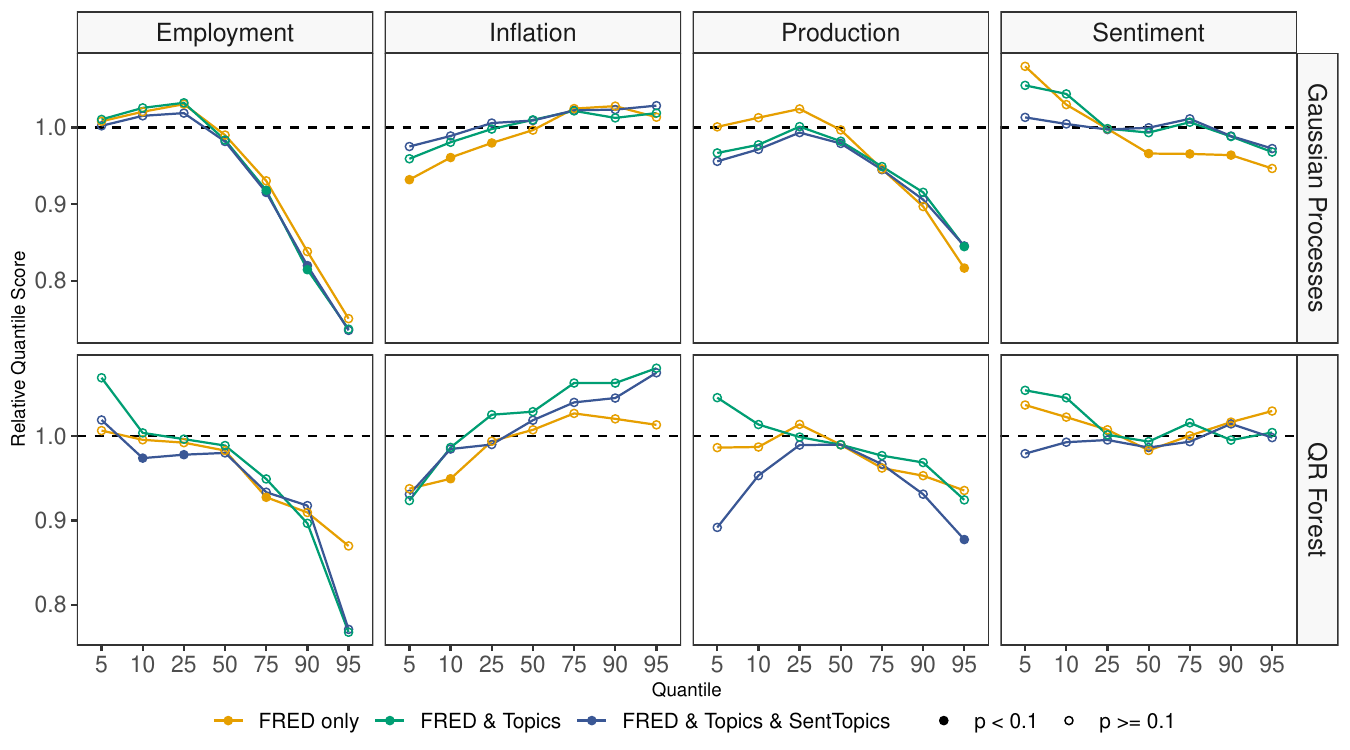}
   \label{fig:Ng2}
\end{subfigure}
\vspace{-.5cm}
\caption{\footnotesize Evaluation of quantile nowcasts (h=0) and quantile one month ahead forecasts (h=1) for Gaussian Processes and QR Forests. The dotted black horizontal line shows the quantile score of Ridge with the full predictor set FRED \& Topics \& SentTopics, which is standardized to 1.0. Scores below (above) 1.0 indicate more (less) precise forecasts for a given quantile compared to the Ridge benchmark. Inside colored dots indicate significantly higher forecast accuracy compared to Ridge according to a one-tailed \cite{diebold1995} test at the $10\%$ level.}
\label{nonlinear}
\end{figure}

Next, we look at the performance of non-linear methods, namely Gaussian Processes and QR Forests. As we are interested in the added value of non-linear models over linear ones, we use Ridge with the full predictor set (FRED \& Topics \& SentTopics) as our benchmark. We do this because Ridge was the overall most successful linear model, particularly in the tails (see Figure \ref{linear}). Figure \ref{nonlinear} shows the relative predictive performance of both non-linear methods with the various predictor sets. Relative quantile scores below (above) one indicate more (less) precise forecasts compared to Ridge with all predictors (FRED \& Topics \& SentTopics). 

As indicated by the predominantly hump-shaped relative quantile scores, with many values below one in Figure \ref{nonlinear}, both non-linear methods were overall better at tail forecasting than Ridge. This is particularly true for the right-tail outcomes of employment and both tails of production, especially when using the FRED \& Topics \& SentTopics predictor set. Hence, the combination of non-linear methods and a mix of economic and text-based predictors seems to be a promising approach for macroeconomic tail forecasting. That said, the marginal difference in quantile scores from adding text-based predictors is smaller overall for non-linear methods than for linear ones. One explanation for this finding may be that text-based predictors may partially compensate for missing non-linear predictive relationships in the linear models.

% We find the following key patterns of predictability for our four target variables. For employment, we observe substantial outperformance in the right tail for now- and forecasts in case of the Gaussian processes; regarding inflation now- and forecasts, accuracy gains are most pronounced in the tails for Gaussian processes and QR forests. Here, Ridge becomes competitive with these methods once textual predictors are added. For industrial production now- and forecasts, overall Gaussian processes exhibit the strongest forecast performance, with QR forests as a close second. Again, the methods with a linear predictive relationship become competitive once textual predictors are added. For consumer sentiment (now- and forecasts), quantile scores markedly improve in the left tail when textual predictors are added, in particular for the methods with a linear predictive relationship. This result is interesting from an information processing aspect, supporting the notion that news data play an important role for households in forming their expectations \citep{larsen2021}.

\color{black}

\subsection{Robustness checks and further analyses}
%\color{blue} 
%In this section we discuss how the value of textual data depends on different choices necessary for the creation of textual predictors and how stable our results are over the evaluation sample. 
We conducted a number of robustness checks and further analyses to explore how alternative settings affect our findings in this paper. 

\textbf{Generating text-based predictors.} The computed topic proportions obtained from the CTM are influenced by a number of choices. Here we explore the impact of alternative choices and examine their potential impact on forecast accuracy. First, we computed $K=68$ topics, which is based on the random search approach by \cite{Mimno2014}. More specifically, we ran their algorithm 100 times and chose the average estimated number of topics across all runs, which was 68. Second, following \cite{adaemmer2020}, who use the CTM and a similar corpus, we computed $K=100$ instead of K=80 topics. The authors show empirically that $K=100$ strikes a good balance between prediction of unseen documents and topic interpretation. Third, the CTM is nested within the more sophisticated structural topic model (STM) by \cite{roberts2016}, which further allows to include covariates. We computed an STM to account for the two different newspapers. Fourth, we estimated our baseline model with 15,000 unique words instead of 10,000 unique words based on the highest tf-idf scores. Fifth, for our reported results, we used a training sample for the CTM from 1980:06 to 1999:08 (see also Section \ref{setup}), and computed OOS topic proportions for the evaluation period from 1999:08 onwards. This approach avoids any look-ahead bias, but implicitly assumes that the vocabulary remained constant over time. In reality, however, this was certainly not the case. For example, words like ``Brexit'' or ``COVID'' were not part of the vocabulary in the estimation sample of the topic proportions. To investigate whether---and if so how much---a fixed vocabulary lowers forecast accuracy, we estimated the CTM using the vocabulary of the whole sample (i.e., with a look-ahead bias).  Interestingly, we did not get lower quantile scores overall. These results are consistent with those of \cite{vandijk2023}, who accounted for shifts in the vocabulary used to nowcast GDP, but did not find gains in forecast accuracy either. Sixth, we computed separate topic proportions for the last week of the month and for each topic. However, we found no added value from this alternative specification based on more recent news. Overall, the robustness checks reveal that the empirical results are not sensitive to the exact specification used to generate the topic proportions. Figure \ref{robtop0} and Figure \ref{robtop1} in Appendix \ref{topalt} summarize the results for the different choices.  %of the generation predictor set, which is in each case 
The predictor set used in each setting is FRED \& Topics \& SentTopics. For comparison, we also added the results for our baseline model ($K=80$). Overall, the results are robust across different choices. We have taken a conservative approach in this paper by avoiding fine-tuning the parameters that control the generation of the text-based data. More sophisticated procedures for selecting these parameters may further increase the gains in macroeconomic quantile prediction, but with an increased risk of overfitting.  

\textbf{Alternative forecast horizons.}
Figure \ref{linear6} and \ref{nonlinear6} in Appendix \ref{hor} summarize the results for the 6-month and the 12-month forecast horizon. In summary, the results confirm our findings for the nowcast and one month ahead forecasts. The marginal improvement from adding (tone-adjusted) text-based predictors seems even higher at these horizons, especially for the non-linear models and for tail predictions associated with economic downside risk (that is, low employment, high inflation and low production).

\textbf{Quantile scores over time.}
Figure \ref{cumloss0} and \ref{cumloss1} in Appendix \ref{qscum} show the cumulative quantile scores at the 10\% quantile relative to the quantile AR(1) benchmark for the nowcasts and one month ahead forecasts, respectively; Figure \ref{cumloss2} and \ref{cumloss3} in Appendix \ref{qscum} do the same for the 90\% quantile. Note that the presentation of results starts at 2005:01 rather than 1999:08, as earlier starting dates would distort the scaling due to the few observations at the beginning of the OOS period. Overall, across the different target variables and methods, gains relative to the quantile AR(1) benchmark are most pronounced during the Great Recession and in the COVID-19 pandemic. Thus, our results are consistent with the typical finding in the economic forecasting literature that the performance of forecasting models is unstable over time; see, e.g. \cite{Rossi2013}. To assess the marginal gains of the text-based predictors, Figure \ref{cumloss0added} and \ref{cumloss1added} in Appendix \ref{qscumadded} show the cumulative quantile scores at the 10\% quantile relative to the FRED-only dataset as the benchmark for the nowcasts and one month ahead forecasts, respectively; Figures \ref{cumloss2added} and \ref{cumloss3added} in Appendix \ref{qscumadded} do the same for the 90\% quantile. Across the different target variables and methods, we observe heterogeneous patterns and, overall, find that text-based predictors (both unadjusted and tone-adjusted) add value relative to the FRED-only dataset at different points in time rather than being restricted to short episodes.  

% \textbf{Outperformance of the non-linear forecasting models against the AR(1) model.}
% Table \ref{outperf} in Appendix \ref{outp} reports the relative share of cases a given non-linear model (either Gaussian Processes or QR Forests), with a given predictor set and a given horizon outperformed the AR(1) benchmark totalled over the 5\%, 10\%, 90\% and 95\% quantiles. Overall, the table indicates that the non-linear models beat the benchmark in most cases.
 %\color{black}\\ % The variation in forecast accuracy is greatest during the Great Recession.\color{black}

\subsection{Which predictors determine the quantile predictions?}

%\color{red}explain more how we measue importants see: https://arxiv.org/pdf/2202.13793.pdf section 3.8 show LASSO regression and cite the paper. see also descriptions for GP regression \color{black}

As we include forecasting methods that feature non-linear predictive relationships, it is not obvious how to determine the marginal effect of a predictor on the variable of interest. Furthermore, even though we use heterogeneous forecasting methods, we want to ensure comparability of measures of predictor importance across methods. To accomplish this task, and to shed light on the predictors with the highest impact, for each forecasting model, we rely on a linear approximation to the prediction distribution for each prediction model, as proposed by \cite{woody2021} and used by \cite{clark2022f} and \cite{Pruser2023}. Concretely, we approximate the quantile predictions $
Q_{\tau ,t+h}$ with a linear regression model that uses regularization.
For each quantile $\tau $, the following Lasso-type optimization problem is
solved:
\begin{equation}
\mathbf{\beta }_{\tau }^{\ast }=\underset{{\beta }_{\tau }}{\arg \min 
}\sum \limits_{t=t_{0}}^{T-h}\left( Q_{\tau ,t+h}-{\beta }_{\tau
}^{\prime }{x}_{t}\right) ^{2}+\lambda
\sum \limits_{j=1}^{J}\left \vert \beta _{\tau ,j}\right \vert \text{,}
\label{lasso}
\end{equation}

where $t_{0}$ denotes the beginning of the hold-out period, ${\beta }
_{\tau }=\left( \beta _{\tau ,1},\ldots,\beta _{\tau ,J}\right) ^{\prime }$
denotes the vector of coefficients, $J$ denotes the number of potential predictors, and $\lambda \geq 0$ is a penalty parameter that controls the shrinkage intensity and is chosen via cross-validation. For the predictor importance analysis we use our most comprehensive set of predictor variables, including both FRED and all text-based predictors. We focus on the $10\%$ and 90\% quantiles because our previous analysis has shown the most pronounced patterns in the tails. 

\begin{figure}[H]
\centering
   % \hspace*{2.38cm}
	%	\includegraphics[width=\textwidth]
    \includegraphics[width=\textwidth]{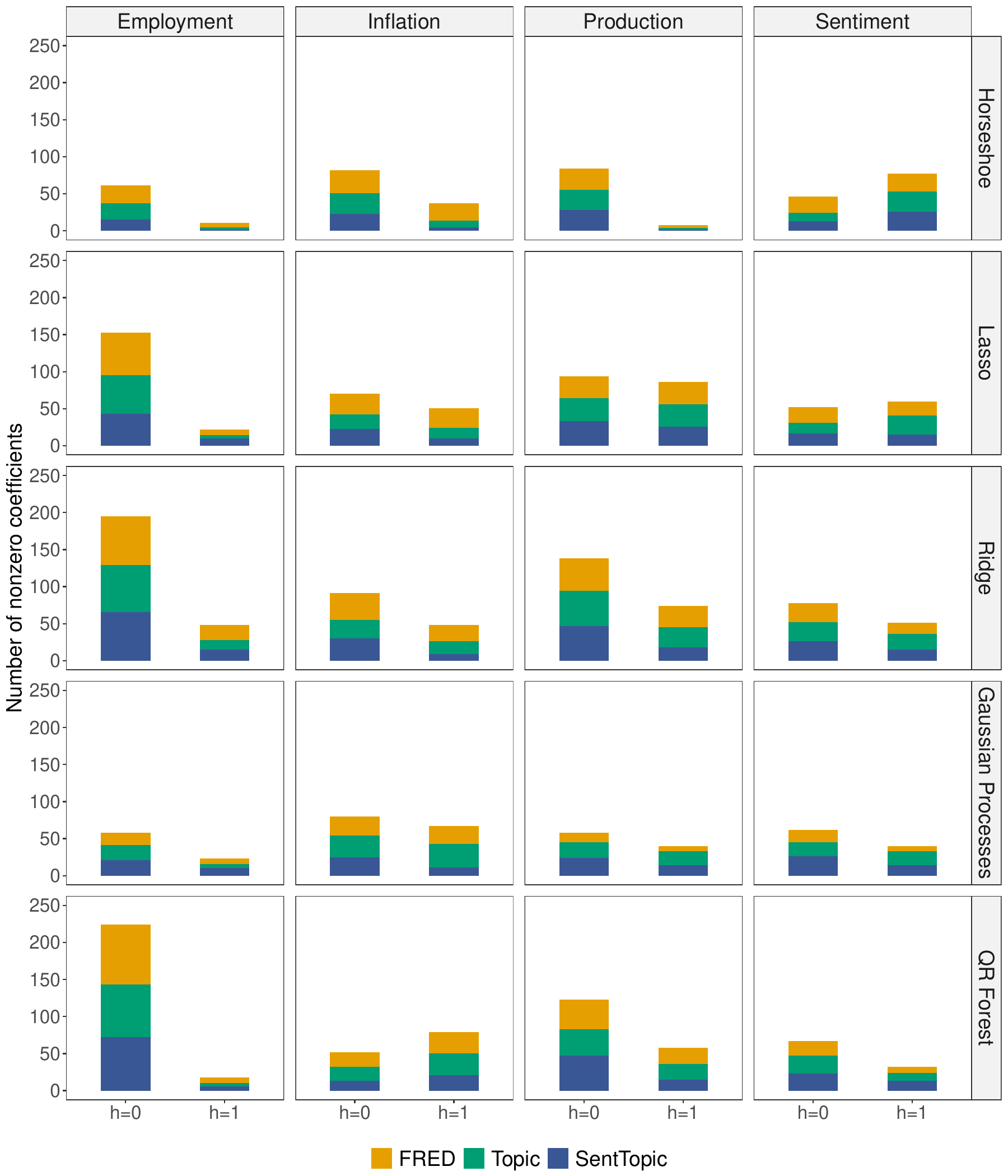} 
				\caption{\footnotesize  Number of non-zero predictors in the regularized regression for the $10\%$-quantile ($\tau=0.1$) for nowcasts (h=0) and one month ahead forecasts (h=1).}
	\label{num_lasso_q10}
\end{figure}

\begin{figure}[H]
\centering
   % \hspace*{2.38cm}
	%	\includegraphics[width=\textwidth]
    \includegraphics[width=\textwidth]{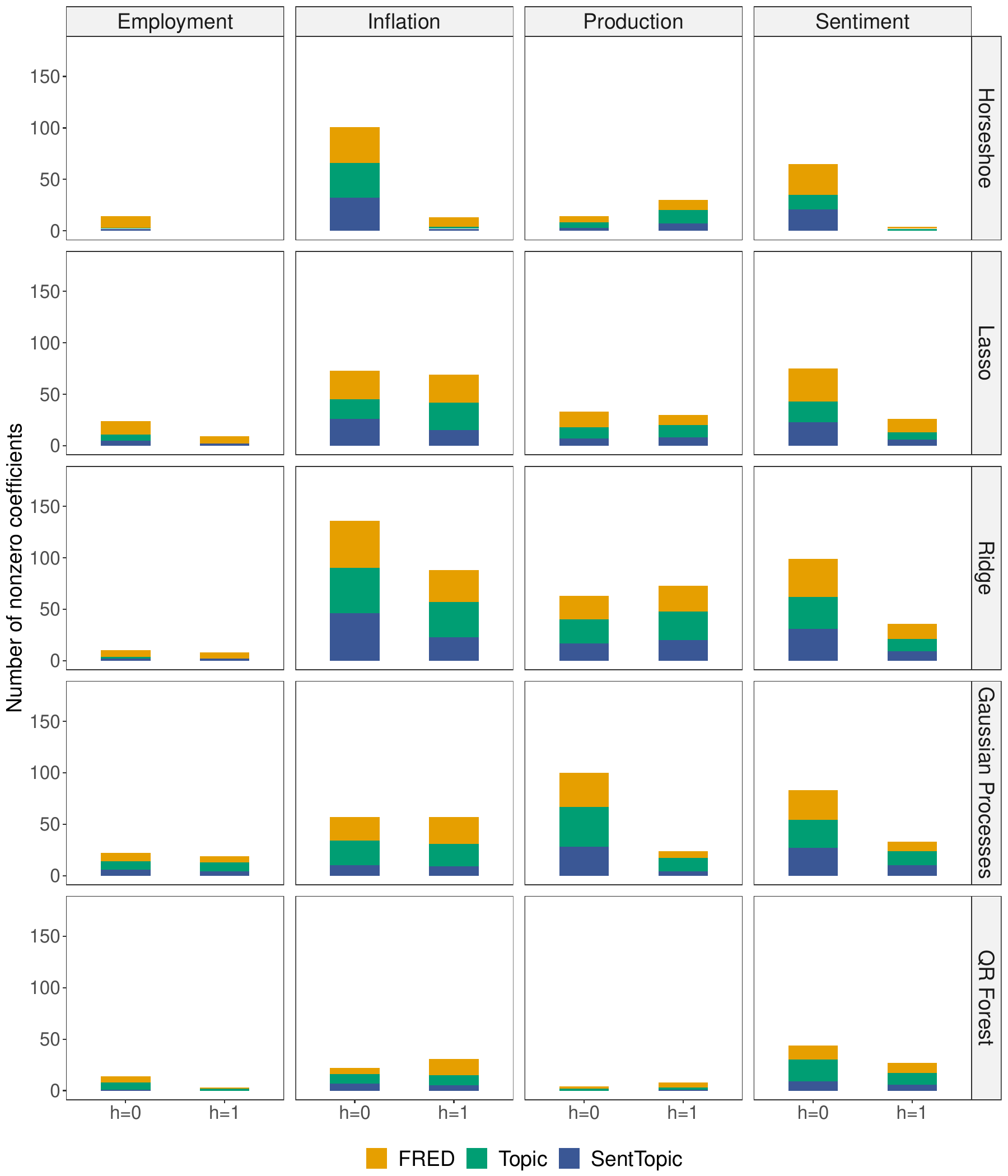} 
				\caption{\footnotesize Number of non-zero predictors in the regularized regression for the $90\%$-quantile ($\tau=0.9$) for nowcasts (h=0) and one month ahead forecasts (h=1).}
	\label{num_lasso_q90}
\end{figure}

%How many \textit{FRED} and textual predictors survived in the Lasso-regression (\ref{lasso})?
Figure \ref{num_lasso_q10} and \ref{num_lasso_q90} show the number of nonzero coefficients for the nowcasts and one month ahead forecasts of the 10\% and the 90\% quantile, respectively. Across different target variables and quantiles, the structure of included FRED and text-based predictors is balanced. Within text-based predictors, roughly the same number of tone-adjusted and unadjusted topic proportions are included. In some cases, however, only text-based predictors survive as in the case for nowcasts of employment with Ridge or QR Forests.

Figure \ref{varimp00} and \ref{varimp11} in Appendix \ref{var_imp_single} report the five most influential predictors  for nowcasts and one month ahead forecasts at the $10\%$-quantile, respectively. Figure \ref{varimp22} and \ref{varimp33} in Appendix \ref{var_imp_single} do the same for the $90\%$-quantile. Importance is measured by the absolute values of the coefficients associated with the (standardized) predictors. 

Overall, the coefficients that are not set to zero, are of different types: lagged values of the respective target variable, FRED economic indicators, unadjusted topic proportions and tone-adjusted topic proportions. Regarding the text-based predictors, the chosen topic proportions make narrative sense for the respective target variable. For instance, both Topic 46 and SentTopic 46 were chosen for inflation nowcasts and one month ahead forecasts. The five most probable words in that topic are related to the housing market  (\textit{home}, \textit{house}, \textit{property}, \textit{estate} and \textit{owner}). Developments in the housing market can obviously be related to inflation dynamics. Topic 31, which is related to household finance, was also chosen for inflation nowcasting by QR Forests. %It  which is plausibly related to inflation and which has the most prominent words \textit{income}, \textit{credit}, \textit{consumer}, \textit{card} and \textit{household}. 
For nowcasts and one month ahead forecasts of employment, Topic 47 and SentTopic 47 were chosen by Gaussian Processes which makes sense, given the topic's job market related words \textit{job}, \textit{work}, \textit{force}, \textit{unemployment} and \textit{employment}. In addition, Topic 55 and SentTopics 55 were frequently selected for nowcasts and one month ahead forecasts of employment and production. The topic is related to nutrition and health issues, given the topic's most prominent words \textit{food}, \textit{disease}, \textit{meat}, \textit{animal} and \textit{product}.

%We have shown in Figure \ref{Sent_figure} how tone-adjustment changes the dynamics of the unadjusted time series.  
\color{black} 

\section{Concluding Remarks}
We have analyzed the added value of text-based predictors for quantile predictions of macroeconomic time series. Our high-dimensional setup included forecasting methods with both linear and non-linear quantile-specific predictive relationships, and we considered different sets of predictors.

Although forecast performance varied across quantiles and target variables, altogether, combinations of FRED and text-based predictors produced the most accurate forecasts, especially in the tails. Tone-adjusted text-based predictors added further value as predictors compared to unadjusted ones alone. Overall, Gaussian Process Regressions and QR Forests prevailed in terms of tail forecast accuracy, suggesting that non-linear predictive relationships are a promising route to follow in tail forecasting. 

In summary, our empirical results suggest that researchers using FRED data for tail risk forecasting should exploit the additional information contained in textual data. We find that the most precise tail risk predictions can be achieved  when using non-linear models with economic indicators and (tone-adjusted) topic proportions. Our results can be of value to researchers, forecasters and policy makers alike. %These results could be of interest to researchers, forecasters and policy makers alike. %These results might be intersting for forecasters and policymakers.  %Based on the findings of our forecasting exercises, using a mix of economic and textual predictors in flexible non-linear methods seems to be the best bet for macroeconomic tail forecasting.\color{black}
%In cases where adding the topic proportions did not provide gains in forecast accuracy, they were not detrimental either. % the differences are negligible. 
%Hence, the benefits from incorporating textual data tend to outweigh the risks.

% In cases where adding the topic proportions did not provide gains in forecast accuracy, they were not detrimental either. Hence, the benefits from incorporating textual data tend to outweigh the risks.

\newpage
%\singlespacing
\spacing{1.00}
\bibliographystyle{econometrica}

%\section*{References}
\addcontentsline{toc}{section}{\refname}% Add references to ToC (and bookmarks)
\bibliography{Sources_Quant}
\newpage

\appendix
\section{Topics and their most probable words}\label{AppA}

\begin{table}[ht]
    \scriptsize
\centering
\begin{adjustbox}{width = \textwidth, center}
\begin{tabular}{cccccccccc}
  \toprule
Topic 1 & Topic 2 & Topic 3 & Topic 4 & Topic 5 & Topic 6 & Topic 7 & Topic 8 & Topic 9 & Topic 10 \\ 
  \midrule
bill & target & committee & sale & dollar & plant & date & country & debt & bond \\ 
  legislation & balance & chairman & unit & currency & factory & start & world & mexico & rate \\ 
  measure & goal & issue & indicator & exchange & production & correction & nation & crisis & yield \\ 
  vote & release & delegate & purchase & yen & auto & delay & government & brazil & note \\ 
  amendment & ratio & convention & auto & mark & industry & schedule & economy & payment & security \\
 \toprule
Topic 11 & Topic 12 & Topic 13 & Topic 14 & Topic 15 & Topic 16 & Topic 17 & Topic 18 & Topic 19 & Topic 20 \\ 
  \midrule
president & member & subc & israel & decline & germany & restaurant & car & america & rate \\ 
  office & group & appropriation & peace & report & france & wine & truck & question & loan \\ 
  director & board & subcommittee & lebanon & figure & britain & bar & vehicle & idea & interest \\ 
  staff & organization & budget & egypt & quarter & west & coffee & bus & problem & mortgage \\ 
  washington & commission & committee & syria & tenth & east & dinner & road & fact & payment \\
  \toprule
  Topic 21 & Topic 22 & Topic 23 & Topic 24 & Topic 25 & Topic 26 & Topic 27 & Topic 28 & Topic 29 & Topic 30 \\ 
  \midrule
police & book & weapon & speech & state & south & price & oil & official & strike \\ 
  violence & newspaper & arm & event & california & black & cent & iraq & charge & protest \\ 
  group & paper & missile & message & jersey & africa & producer & iran & investigation & movement \\ 
  death & magazine & korea & crowd & governor & minority & gasoline & barrel & case & leader \\ 
  attack & news & defense & address & texas & white & demand & kuwait & trial & government \\
  \toprule
  Topic 31 & Topic 32 & Topic 33 & Topic 34 & Topic 35 & Topic 36 & Topic 37 & Topic 38 & Topic 39 & Topic 40 \\ 
  \midrule
income & show & tax & stock & child & trade & worker & church & gold & hotel \\ 
  credit & los & budget & point & family & japan & union & nicaragua & dollar & room \\ 
  consumer & angeles & cut & investor & life & export & labor & aid & york & island \\ 
  card & san & taxis & index & parent & import & employee & salvador & trading & tourist \\ 
  household & film & spending & share & mother & product & wage & abortion & rate & town \\ 
  \toprule
  Topic 41 & Topic 42 & Topic 43 & Topic 44 & Topic 45 & Topic 46 & Topic 47 & Topic 48 & Topic 49 & Topic 50 \\ 
  \midrule
building & computer & campaign & drug & policy & home & job & city & study & bank \\ 
  project & technology & candidate & crime & official & house & work & york & report & loan \\ 
  housing & system & voter & police & administration & property & force & resident & number & banking \\ 
  construction & information & election & prison & effort & estate & unemployment & neighborhood & research & institution \\ 
  development & software & poll & officer & action & owner & employment & community & survey & saving \\ 
   \toprule
   Topic 51 & Topic 52 & Topic 53 & Topic 54 & Topic 55 & Topic 56 & Topic 57 & Topic 58 & Topic 59 & Topic 60 \\ 
  \midrule
united & plane & china & agreement & food & party & share & program & health & rate \\ 
  states & ship & kong & talk & disease & election & quarter & government & care & economy \\ 
  washington & space & hong & negotiation & meat & leader & profit & agency & insurance & inflation \\ 
  cuba & aircraft & beijing & side & animal & government & earning & aid & hospital & growth \\ 
  official & port & taiwan & deal & product & power & loss & welfare & doctor & interest \\ 
     \toprule
     Topic 61 & Topic 62 & Topic 63 & Topic 64 & Topic 65 & Topic 66 & Topic 67 & Topic 68 & Topic 69 & Topic 70 \\ 
  \midrule
woman & school & service & store & war & market & company & plant & fund & business \\ 
  man & student & airline & chain & force & future & firm & steel & money & industry \\ 
  friend & education & airport & customer & troop & firm & executive & energy & investment & economy \\ 
  life & college & flight & retailer & army & option & deal & power & investor & sector \\ 
  guy & teacher & fare & shopping & military & broker & contract & utility & return & competition \\ 
   \toprule
   Topic 71 & Topic 72 & Topic 73 & Topic 74 & Topic 75 & Topic 76 & Topic 77 & Topic 78 & Topic 79 & Topic 80 \\ 
  \midrule
county & percent & union & vietnam & building & farm & water & meeting & law & control \\ 
  area & increase & soviet & refugee & room & farmer & land & conference & case & thousand \\ 
  maryland & percentage & russia & immigrant & wall & crop & mile & leader & court & million \\ 
  official & average & moscow & immigration & floor & grain & area & statement & lawyer & hundred \\ 
  virginia & rise & poland & border & window & land & tree & official & rule & effort \\ 
  \bottomrule
\end{tabular}
\end{adjustbox}
\end{table}
\newpage

\section{Robustness: Text-based predictors}\label{topalt}
\begin{figure}[H]
\centering
%\vspace{-2cm}
%\renewcommand{\thesubfigure}{Nowcast, h = 0}
\begin{subfigure}[b]{.67\textwidth}
    \subcaption[short for lof]{(Nowcast, h = 0)}
   \includegraphics[width=1\linewidth]{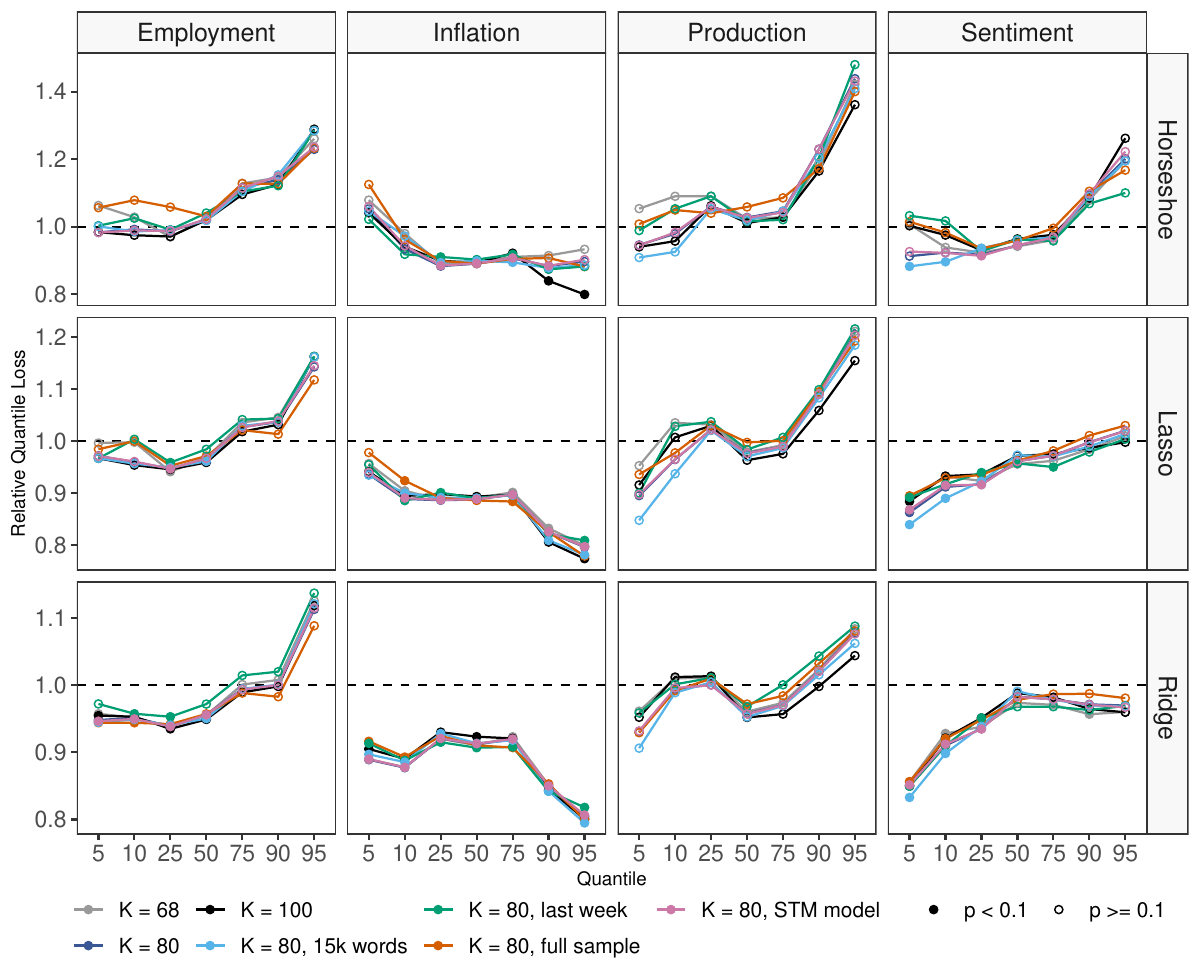}   
   \label{fig:Ng1} 
\end{subfigure}
\begin{subfigure}[b]{.67\textwidth}
\subcaption[short for lof]{(Forecast, h = 1)}
   \includegraphics[width=1\linewidth]{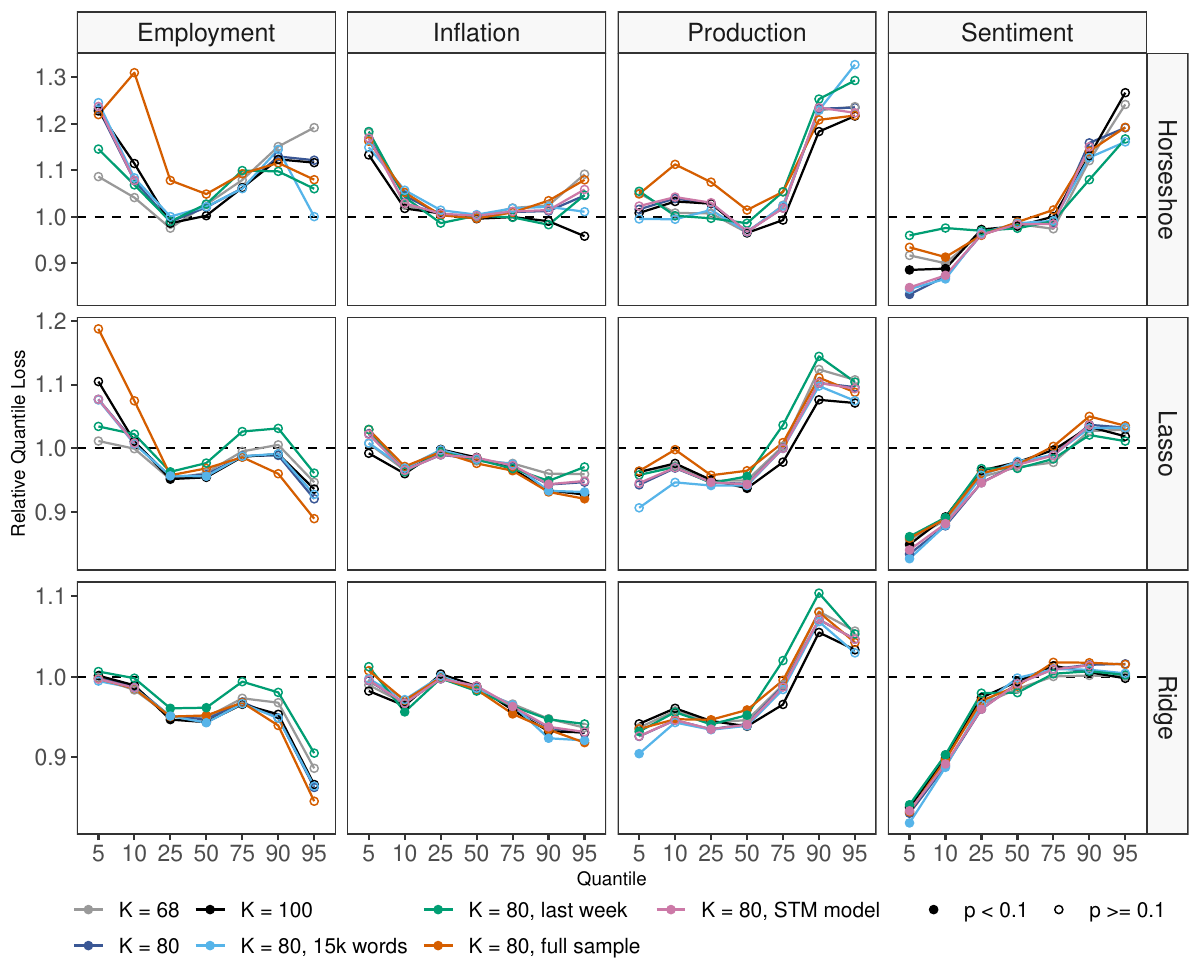}
   \label{fig:Ng2}
\end{subfigure}
\vspace{-.5cm}
\caption{\footnotesize Robustness analyses for alternative choices that affect the generation of the text-based predictors. Each predictor set consists of FRED \& Topics \& SentTopics. K refers to the number of topics. The darkblue line refers to our baseline model (K=80, CTM, 10k unique words, out-of-sample computed topic proportions). The dotted black horizontal line shows the quantile score of the AR(1) benchmark which is standardized to 1.0. Scores below (above) 1.0 indicate more (less) precise forecasts for a given quantile compared to the AR(1) benchmark. Inside colored dots indicate significantly higher forecast accuracy compared to the AR(1) benchmark according to a one-tailed \cite{diebold1995} test at the $10\%$ level. }
% \caption[]{\scriptsize Linear models. One-step-ahead (h = 1) quantile scores.}
\label{robtop0}
\end{figure}

\begin{figure}[H]
\centering
%\vspace{-2cm}
%\renewcommand{\thesubfigure}{Nowcast, h = 0}
\begin{subfigure}[b]{.8\textwidth}
    \subcaption[short for lof]{(Nowcast, h = 0)}
   \includegraphics[width=1\linewidth]{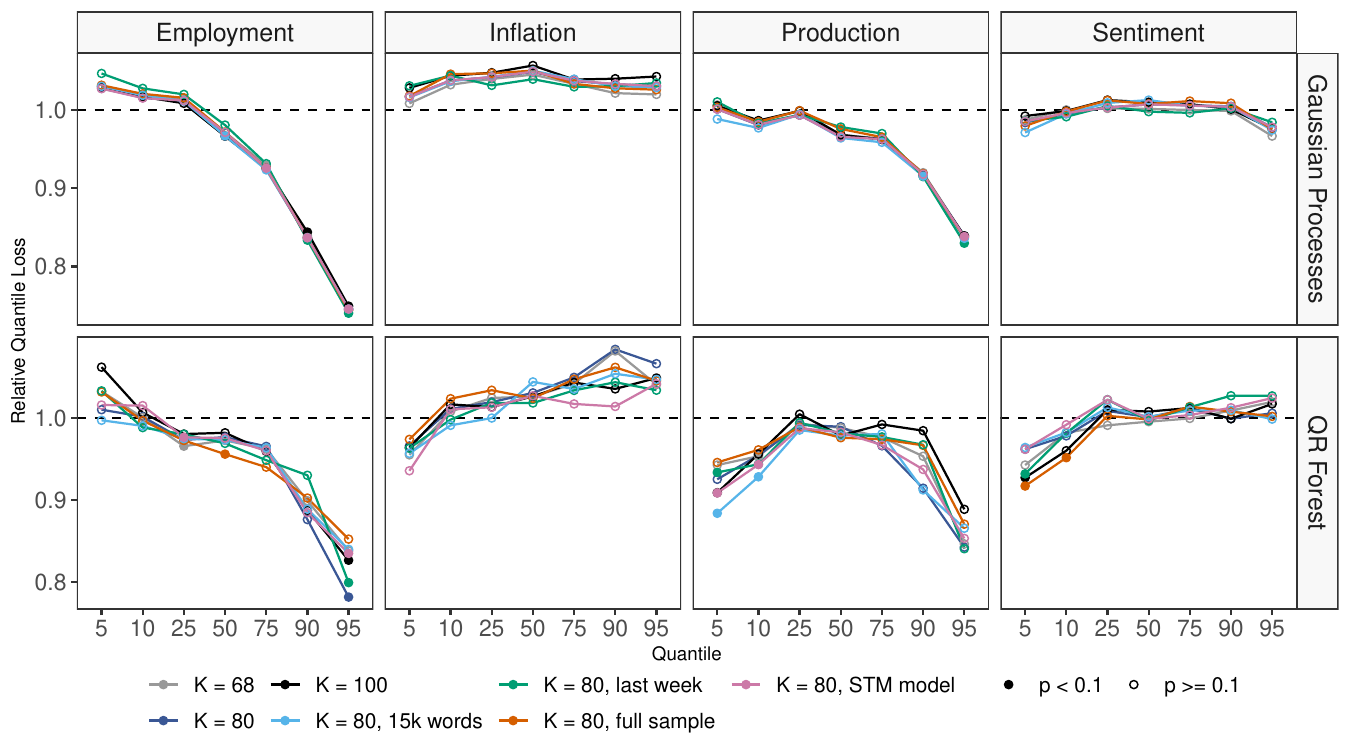}   
   \label{fig:Ng1} 
\end{subfigure}
\begin{subfigure}[b]{.8\textwidth}
\subcaption[short for lof]{(Forecast, h = 1)}
   \includegraphics[width=1\linewidth]{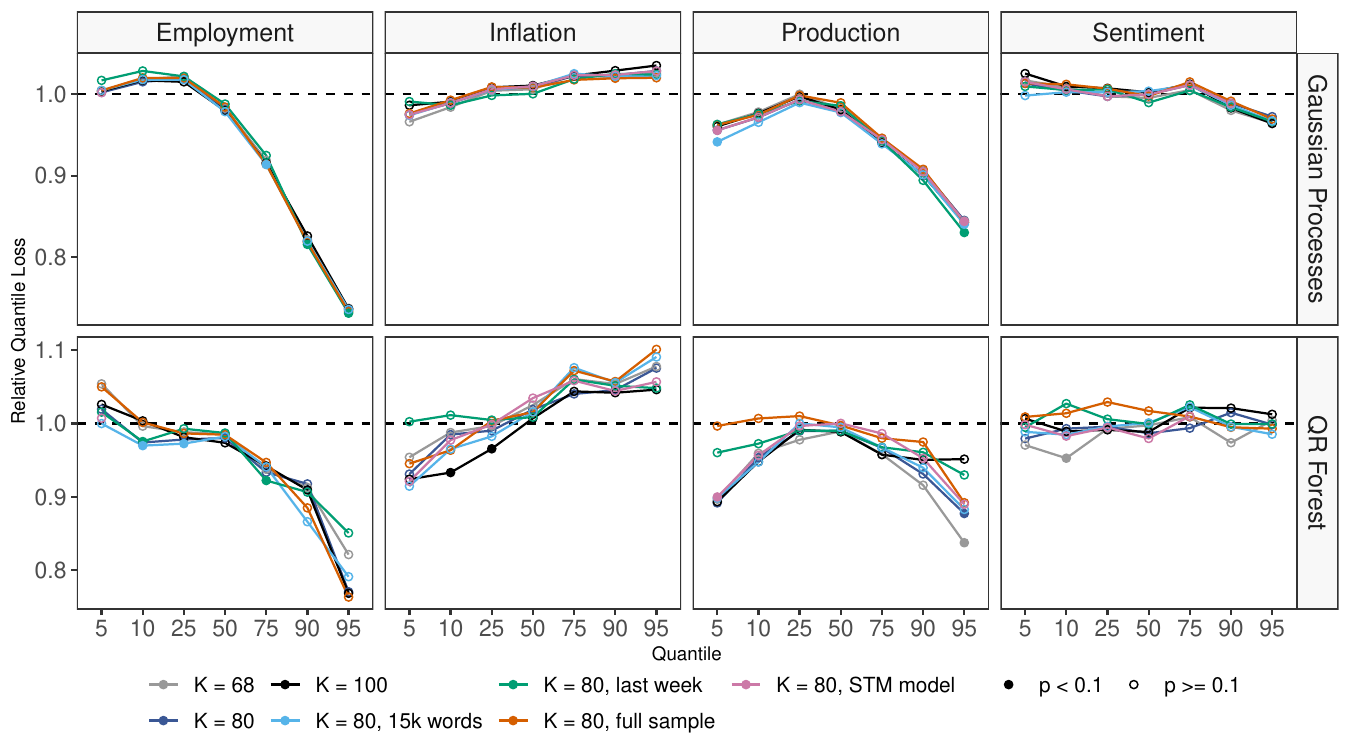}
   \label{fig:Ng2}
\end{subfigure}
\vspace{-.5cm}
\caption{\footnotesize Robustness analyses for alternative choices that affect the generation of the text-based predictors. Each predictor set consists of FRED \& Topics \& SentTopics. K refers to the number of topics. The darkblue line refers to our baseline model (K=80, CTM, 10k unique words, out-of-sample computed topic proportions). The dotted black horizontal line shows the quantile score of the Ridge benchmark with the predictor set FRED \& Topics \& SentTopics, which is standardized to 1.0. Scores below (above) 1.0 indicate more (less) precise forecasts for a given quantile compared to the quantile AR(1) benchmark. Inside colored dots indicate significantly higher forecast accuracy compared to the quantile AR(1) benchmark according to a one-tailed \cite{diebold1995} test at the $10\%$ level.}
\label{robtop1}
\end{figure}

\section{Alternative forecast horizons}\label{hor}

\begin{figure}[H]
\centering
%\vspace{-2cm}
%\renewcommand{\thesubfigure}{Nowcast, h = 0}
\begin{subfigure}[b]{.71\textwidth}
    \subcaption[short for lof]{(Forecast, h = 6)}
   \includegraphics[width=1\linewidth]{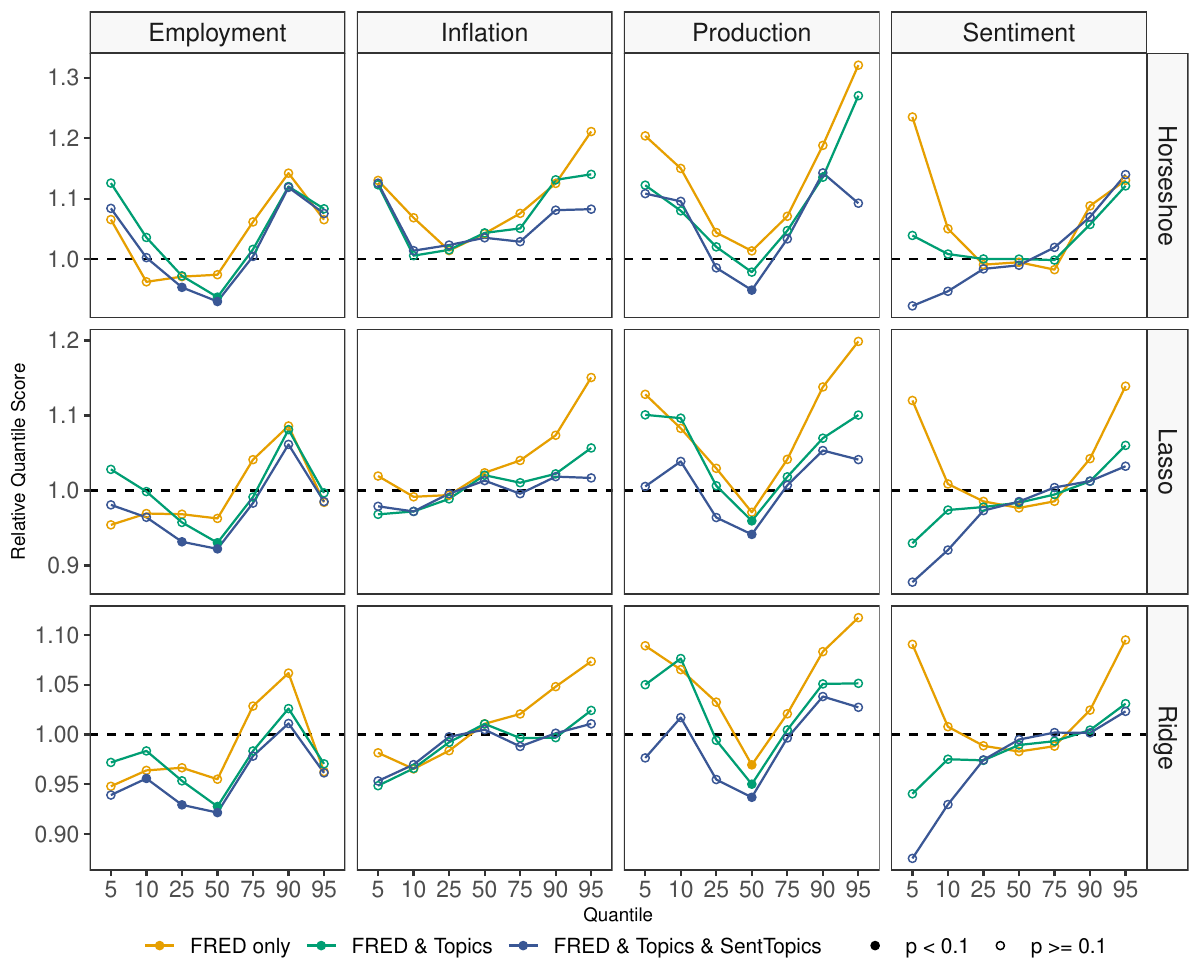}   
   \label{fig:Ng1} 
\end{subfigure}
\begin{subfigure}[b]{.71\textwidth}
\subcaption[short for lof]{(Forecast, h = 12)}
   \includegraphics[width=1\linewidth]{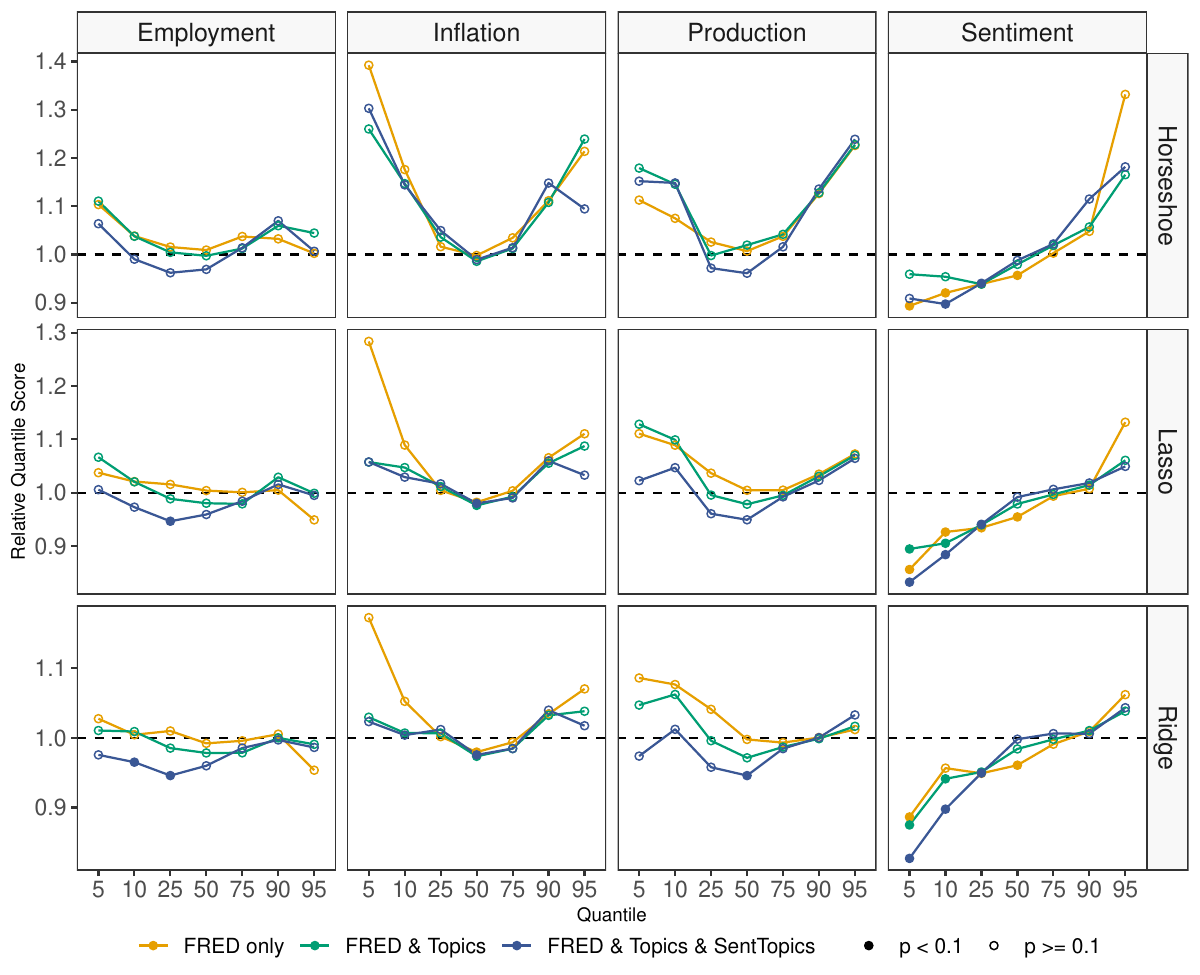}
   \label{fig:Ng2}
\end{subfigure}
\vspace{-.5cm}
\caption{\footnotesize Evaluation of quantile forecasts for h=6 and h=12 for the Bayesian QRs with different shrinkage priors (Horseshoe, Ridge and Lasso). The dotted black horizontal line shows the quantile score of the quantile AR(1) benchmark which is standardized to 1.0. Scores below (above) 1.0 indicate more (less) precise forecasts for a given quantile compared to the quantile AR(1) benchmark. Inside colored dots indicate significantly higher forecast accuracy compared to the quantile AR(1) benchmark according to a one-tailed \cite{diebold1995} test at the $10\%$ level.}
% \caption[]{\scriptsize Linear models. One-step-ahead (h = 1) quantile scores.}
\label{linear6}
\end{figure}

\begin{figure}[H]
\centering
%\vspace{-2cm}
%\renewcommand{\thesubfigure}{Nowcast, h = 0}
\begin{subfigure}[b]{.8\textwidth}
    \subcaption[short for lof]{(Forecast, h = 6)}
   \includegraphics[width=1\linewidth]{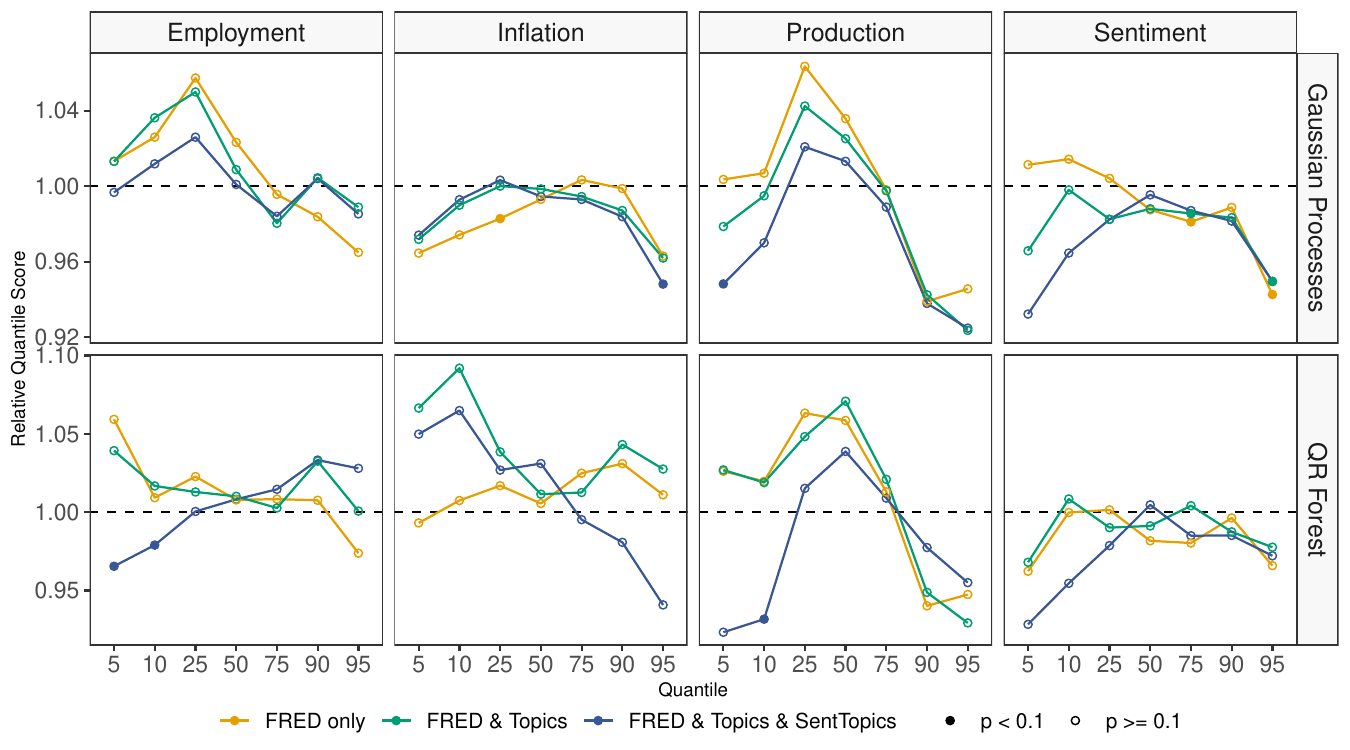}   
   \label{fig:Ng1} 
\end{subfigure}
\begin{subfigure}[b]{.8\textwidth}
\subcaption[short for lof]{(Forecast, h = 12)}
   \includegraphics[width=1\linewidth]{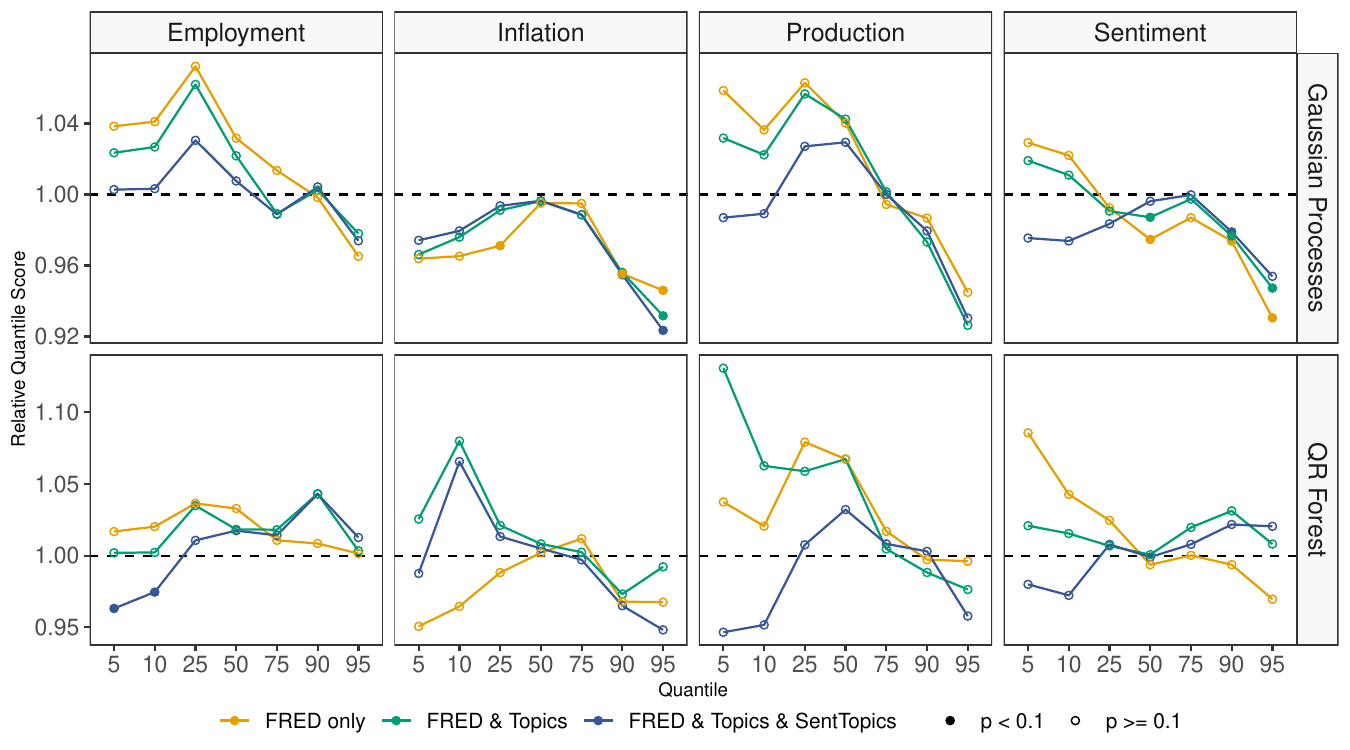}
   \label{fig:Ng2}
\end{subfigure}
\vspace{-.5cm}
\caption{\footnotesize  Evaluation of quantile forecasts for h=6 and h=12 for the non-linear models, QR Forests and Gaussian Processes. The dotted black horizontal line shows the quantile score of the Ridge benchmark with the predictor set FRED \& Topics \& SentTopics, which is standardized to 1.0. Scores below (above) 1.0 indicate more (less) precise forecasts for a given quantile compared to the Ridge benchmark. Inside colored dots indicate significantly higher forecast accuracy compared to the Ridge benchmark according to a one-tailed \cite{diebold1995} test at the $10\%$ level.}
% \caption[]{\scriptsize Linear models. One-step-ahead (h = 1) quantile scores.}
\label{nonlinear6}
\end{figure}

\section{Variable importance}\label{var_imp_single}

\begin{figure}[H]
\centering
		\includegraphics[width=\textwidth]{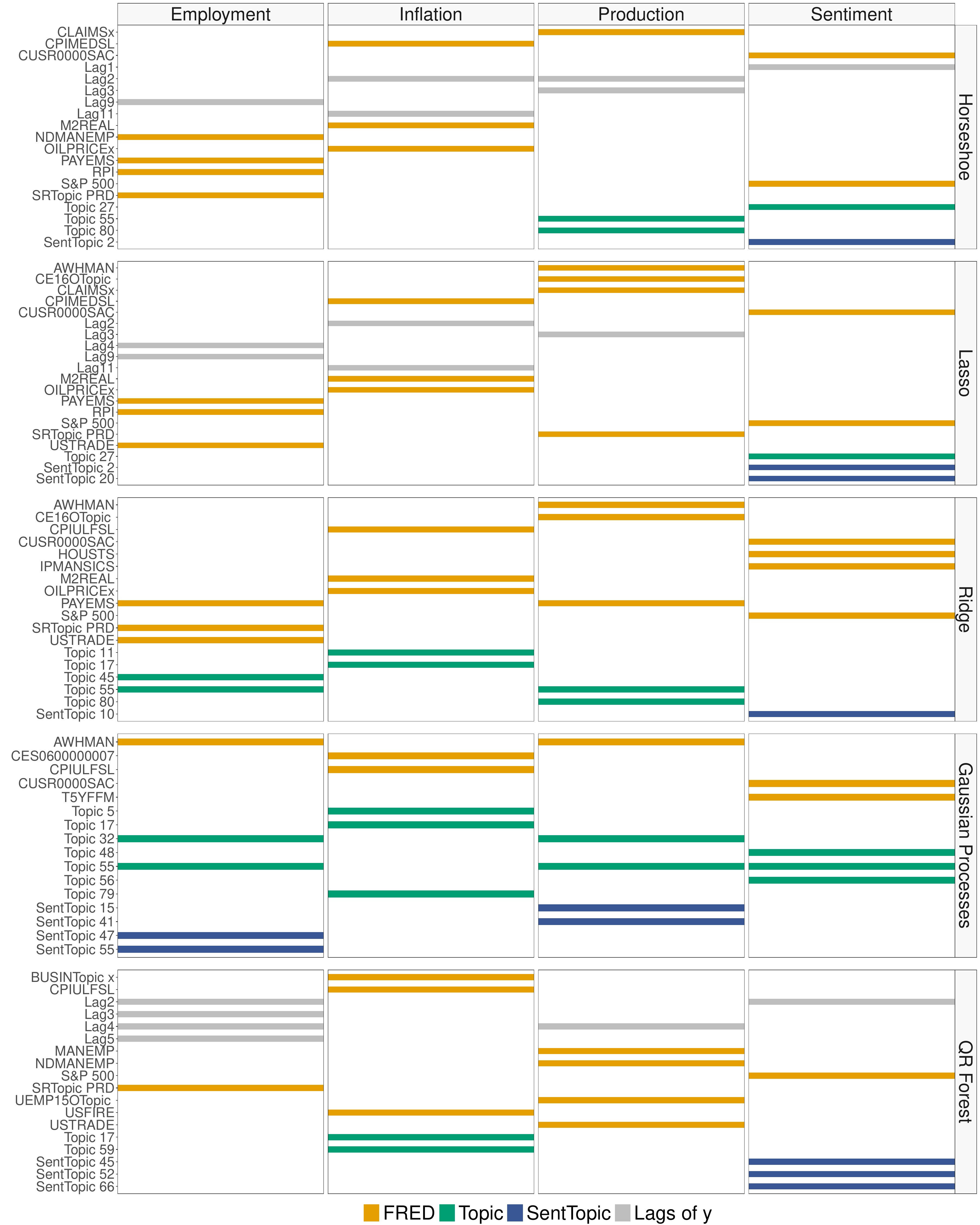} 
			\caption{\footnotesize Variable importance for nowcasts at the $10\%$-quantile $(h=0$, $\tau = 10\%)$.}
		\label{varimp00}
\end{figure}

\begin{figure}[H]
\centering
    \includegraphics[width=\textwidth]{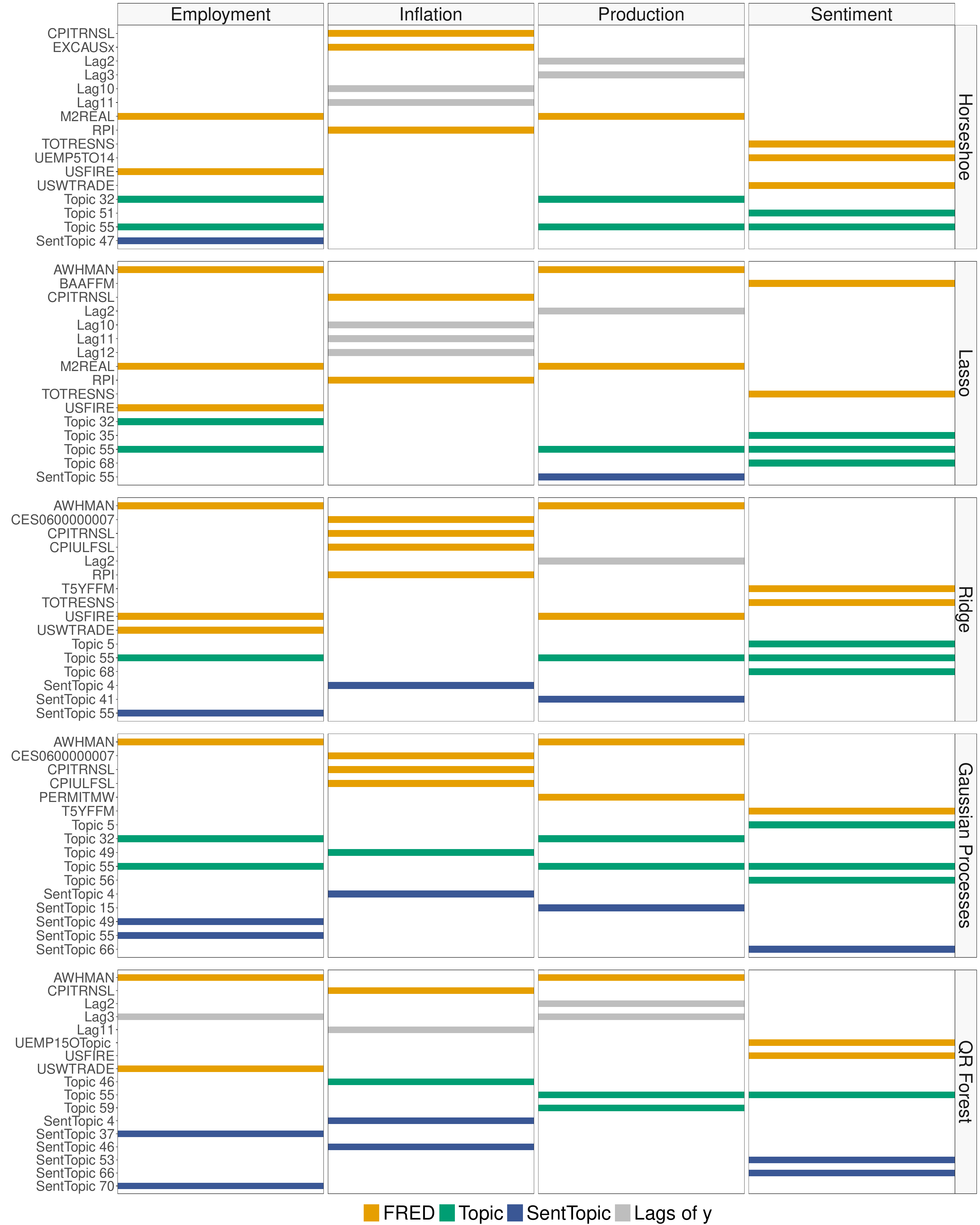} 
				\caption{\footnotesize Variable importance for one month ahead forecasts at the $10\%$-quantile $(h=1, \tau = 10\%)$.}
	\label{varimp11}
\end{figure}

\begin{figure}[H]
\centering
		\includegraphics[width=\textwidth]{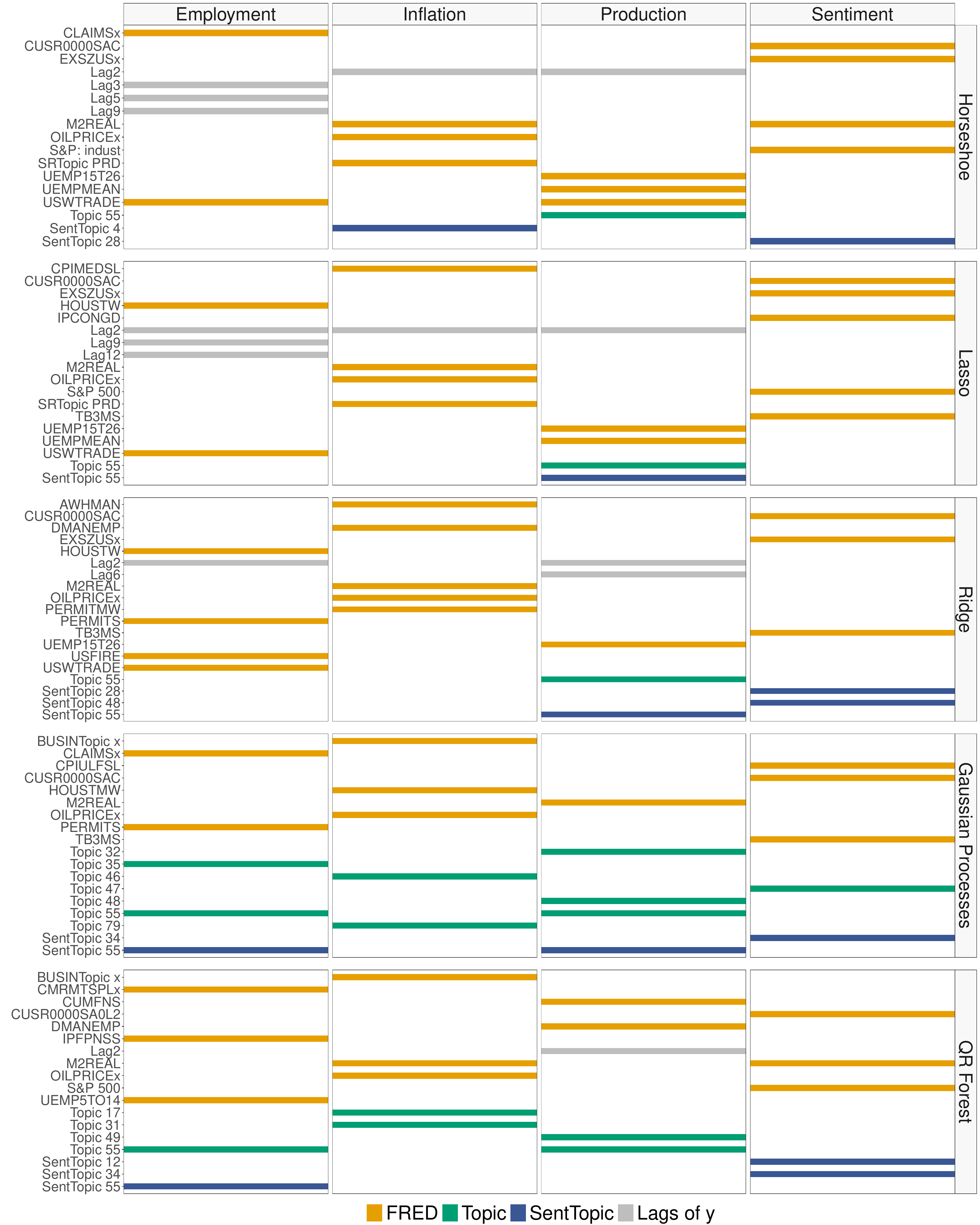} 
			\caption{\footnotesize Variable importance for nowcasts at the $90\%$-quantile $(h=0$, $\tau = 90\%)$.}
		\label{varimp22}
\end{figure}

\begin{figure}[H]
\centering
		\includegraphics[width=\textwidth]{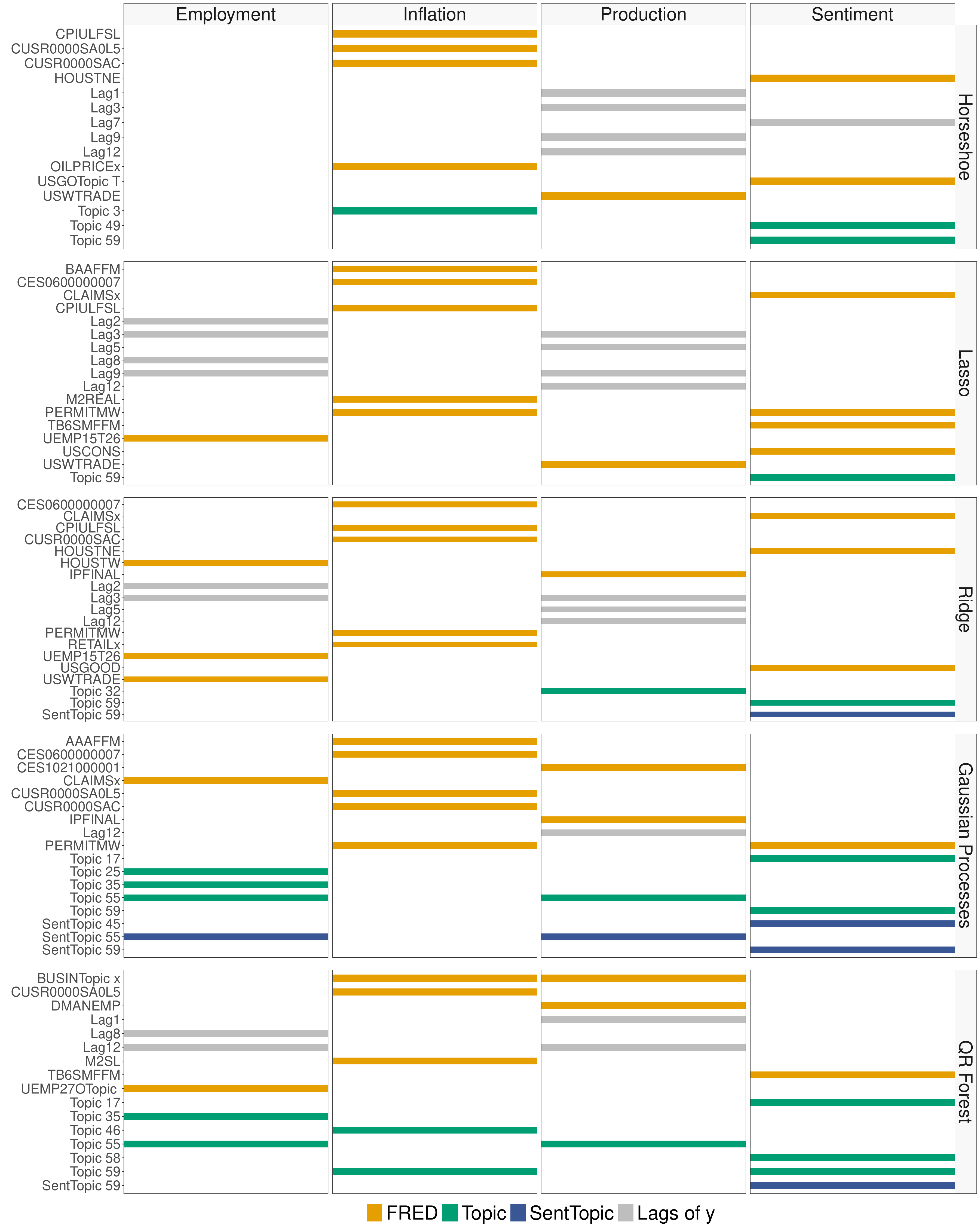} 
			\caption{\footnotesize Variable importance for one month ahead at the $90\%$-quantile $(h=1$, $\tau = 90\%)$.}
		\label{varimp33}
\end{figure}

\section{Quantile scores over time}\label{qscum}

\begin{figure}[H]
\centering
		\includegraphics[width=\textwidth]{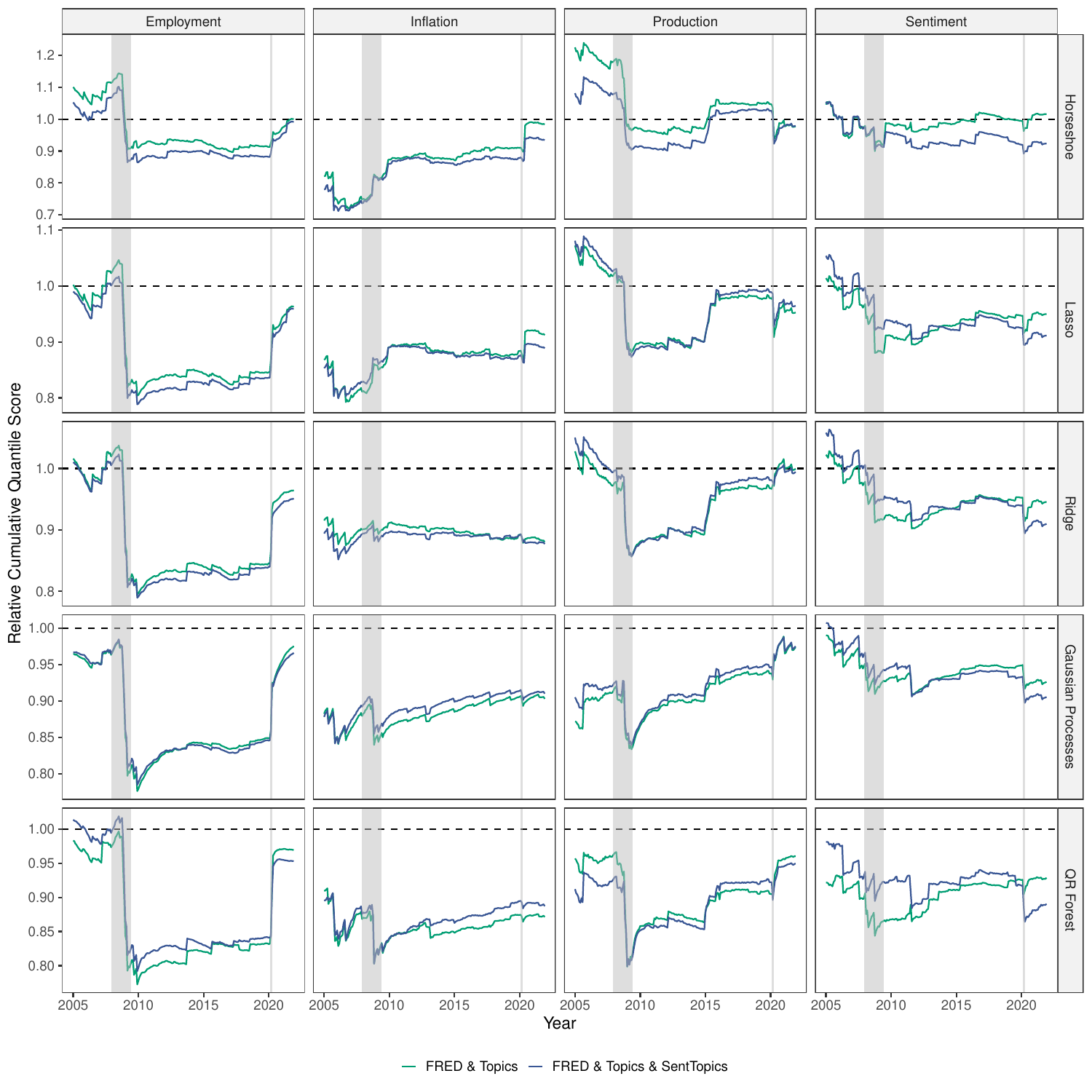} 
		\caption{\footnotesize Nowcast (h = 0) relative cumulative quantile scores for the $10\%$-quantile ($\tau=0.1$) relative to a quantile AR(1) model.   The cumulative quantile scores of the respective benchmarks are standardized to 1.0.  Scores below (above) 1.0 indicate more (less) precise forecasts compared to the respective benchmark until the given point in time. Gray-shaded areas indicate NBER-dated recessions.  }
		\label{cumloss0}
\end{figure}

\begin{figure}[H]
\centering
		\includegraphics[width=\textwidth]{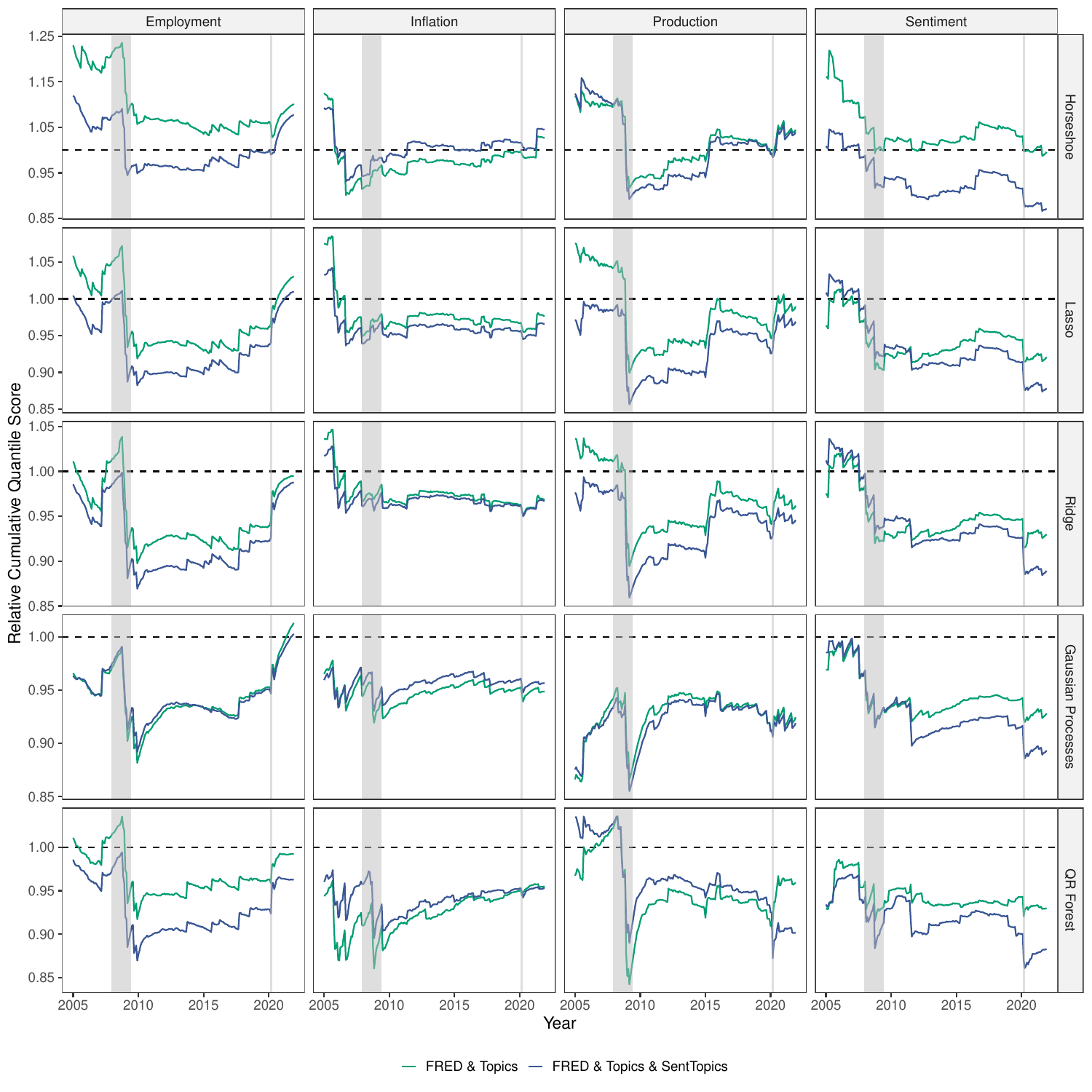} 
		\caption{\footnotesize One month ahead (h = 1) relative cumulative quantile score for the $10\%$-quantile ($\tau=0.1$) relative to a quantile AR(1) model.    The cumulative quantile scores of the respective benchmarks are standardized to 1.0.  Scores below (above) 1.0 indicate more (less) precise forecasts compared to the respective benchmark until the given point in time.   Gray-shaded areas indicate NBER-dated recessions.  }
		\label{cumloss1}
\end{figure}

\begin{figure}[H]
\centering
		\includegraphics[width=\textwidth]{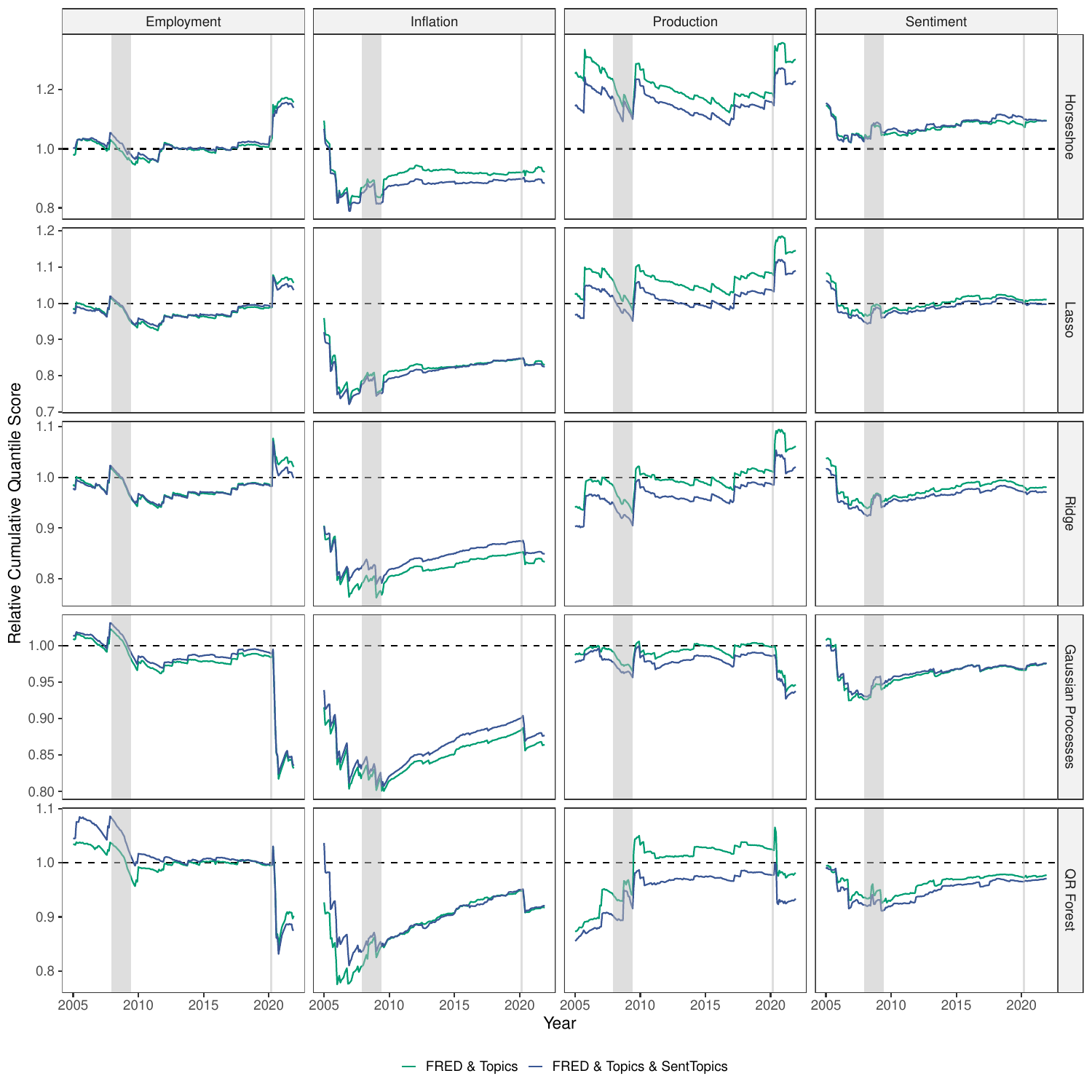} 
		\caption{\footnotesize Nowcast (h = 0)  relative  cumulative quantile score for the $90\%$-quantile ($\tau=0.9$) relative to a quantile AR(1) model.  The cumulative quantile scores of the respective benchmarks are standardized to 1.0.   Scores below (above) 1.0 indicate more (less) precise forecasts compared to respective benchmark until the given point in time. Gray-shaded areas indicate NBER-dated recessions.}
		\label{cumloss2}
\end{figure}

\begin{figure}[H]
\centering
		\includegraphics[width=\textwidth]{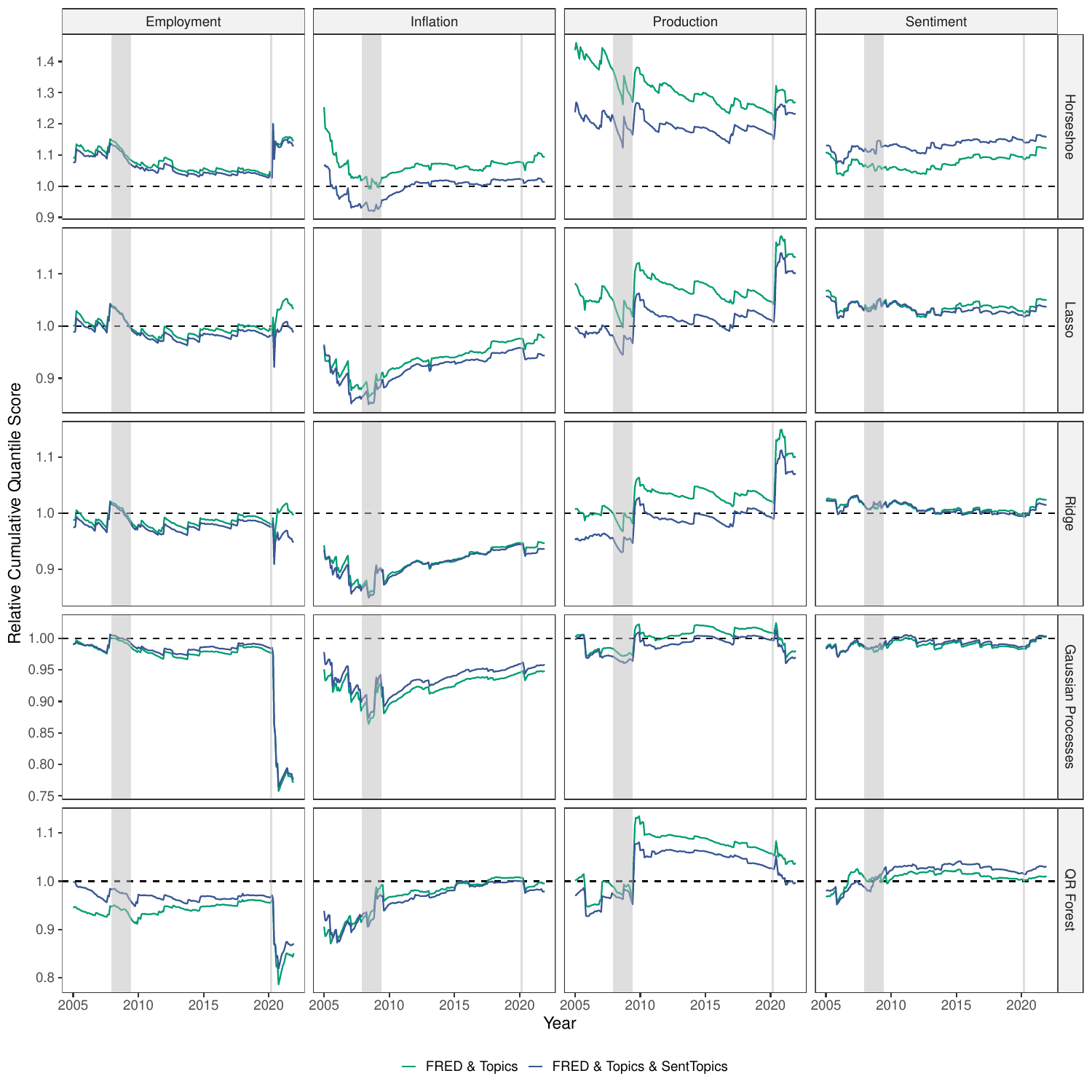} 
		\caption{\footnotesize One month ahead (h = 1) relative cumulative quantile score for the $90\%$-quantile ($\tau=0.9$) relative to a quantile AR(1) model.  The cumulative quantile scores of the respective benchmarks are standardized to 1.0.   Scores below (above) 1.0 indicate more (less) precise forecasts compared to respective benchmark until the given point in time.  Gray-shaded areas indicate NBER-dated recessions.  }
		\label{cumloss3}
\end{figure}

\section{Added value of textual data over time}\label{qscumadded}

\begin{figure}[H]
\centering
		\includegraphics[width=\textwidth]{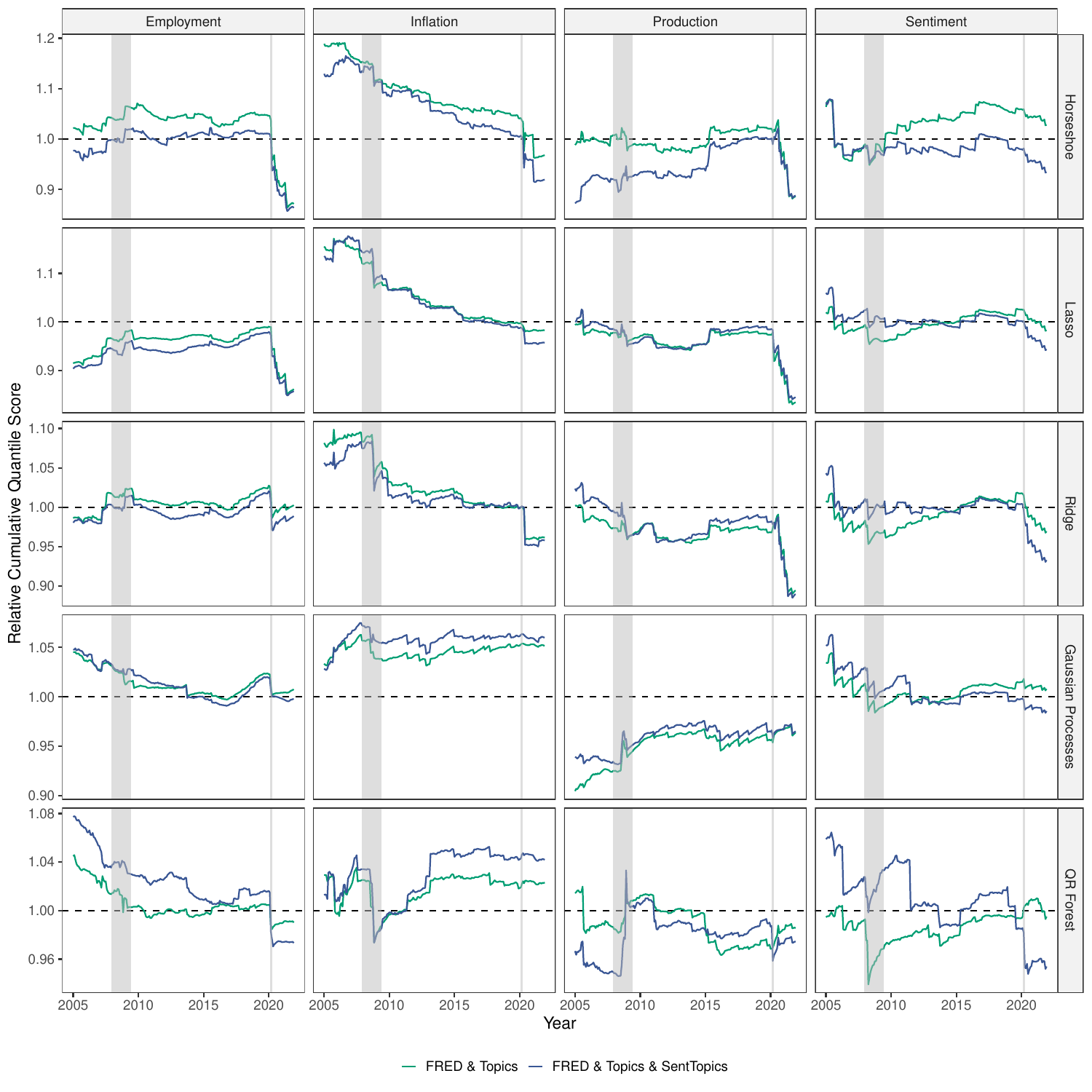} 
		\caption{\footnotesize Nowcast (h = 0)  relative cumulative quantile scores for the $10\%$-quantile ($\tau=0.1$) relative to the respective model and the FRED only benchmark.   The cumulative quantile scores of the respective benchmarks are standardized to 1.0. Scores below (above) 1.0 indicate more (less) precise forecasts compared to the benchmark until the given point in time. Gray-shaded areas indicate NBER-dated recessions.}
		\label{cumloss0added}
\end{figure}

\begin{figure}[H]
\centering
		\includegraphics[width=\textwidth]{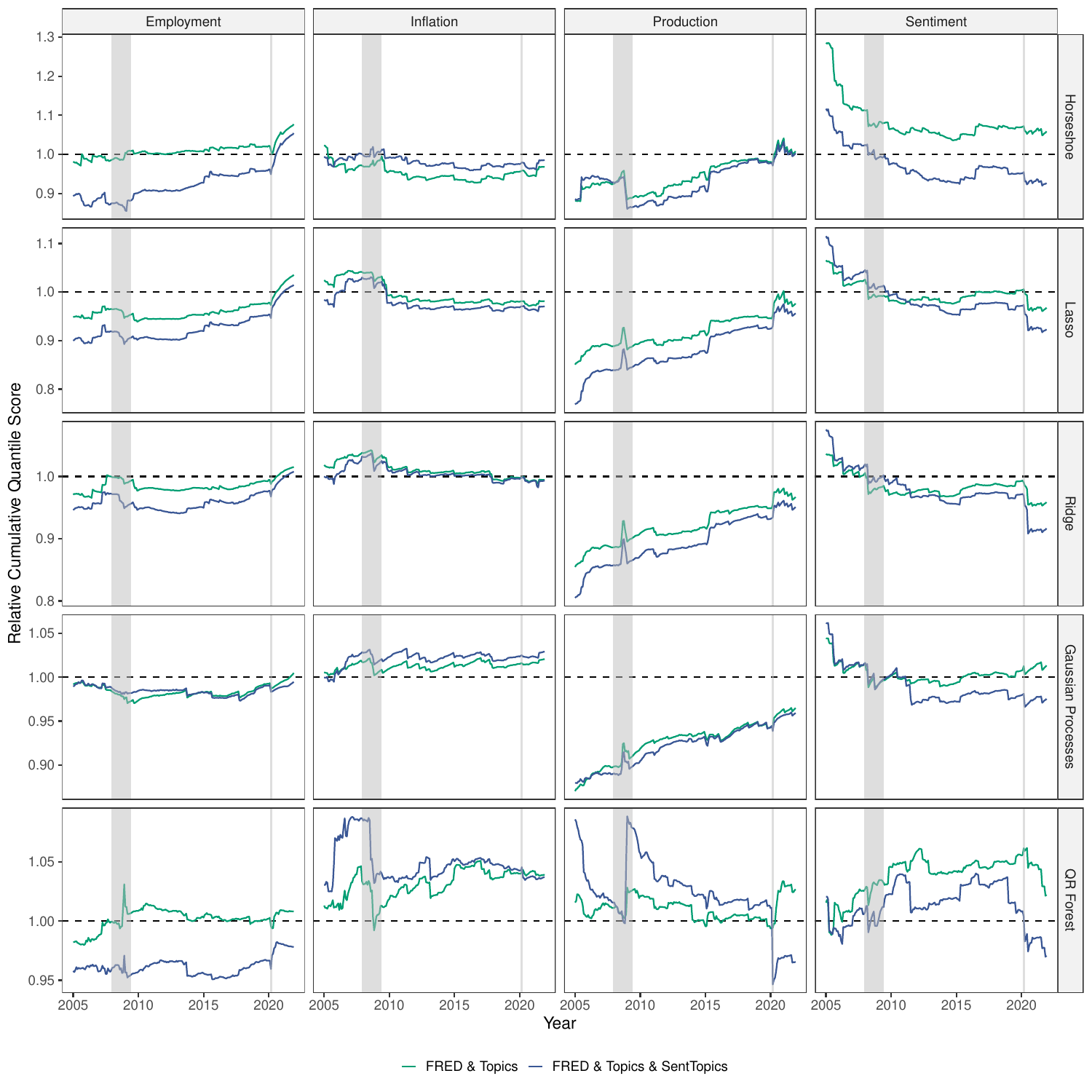} 
		\caption{\footnotesize One month ahead (h = 1)  relative cumulative quantile score for the $10\%$-quantile ($\tau=0.1$) relative to the respective model and the FRED only benchmark.   The cumulative quantile scores of the respective benchmarks are standardized to 1.0. Scores below (above) 1.0 indicate more (less) precise forecasts compared to the benchmark until the given point in time. Gray-shaded areas indicate NBER-dated recessions. }
		\label{cumloss1added}
\end{figure}

\begin{figure}[H]
\centering
		\includegraphics[width=\textwidth]{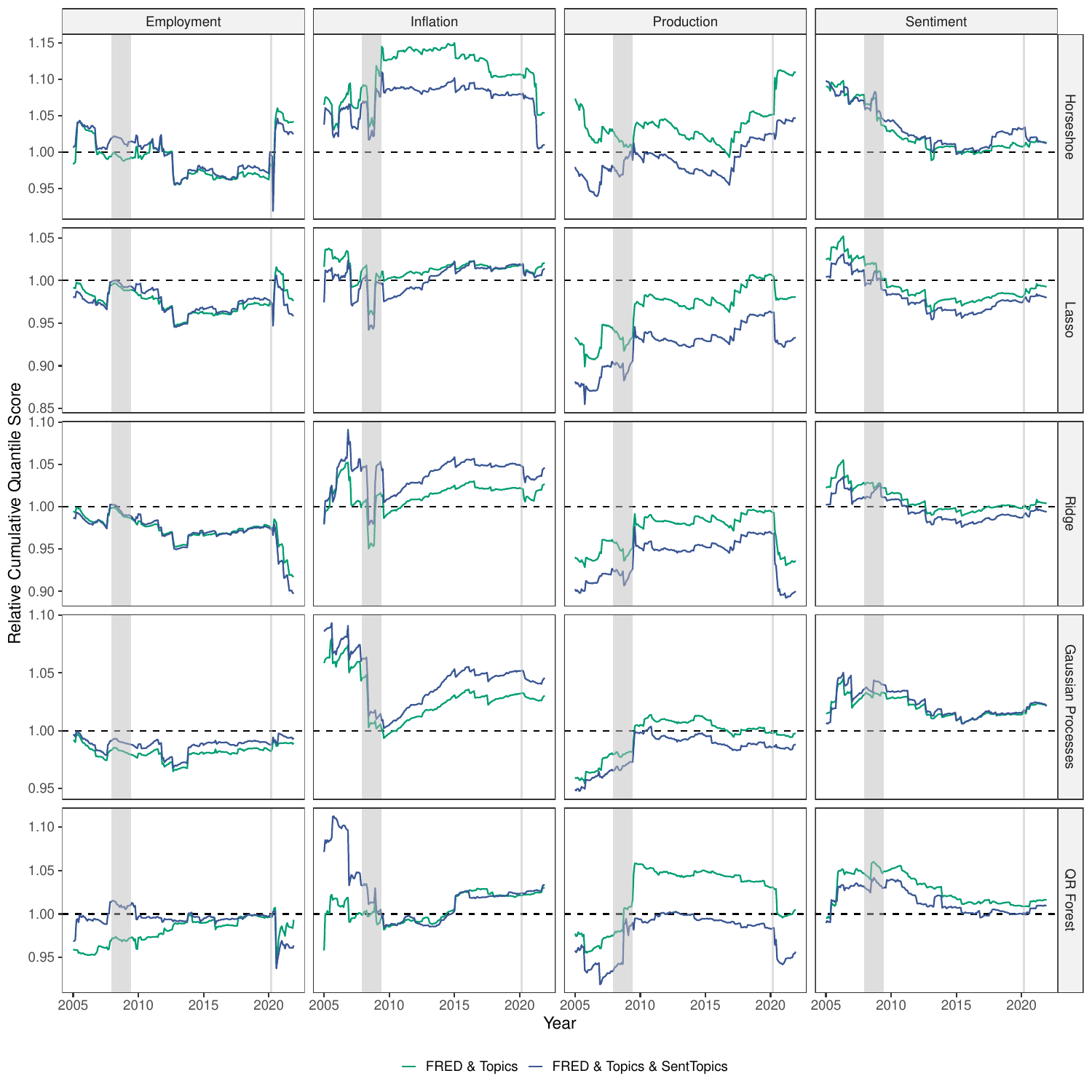} 
		\caption{\footnotesize Nowcast (h = 0) relative cumulative quantile score for the $90\%$-quantile ($\tau=0.9$) relative to the respective model and the FRED only benchmark.   The cumulative quantile scores of the respective benchmarks are standardized to 1.0. Scores below (above) 1.0 indicate more (less) precise forecasts compared to the benchmark until the given point in time. Gray-shaded areas indicate NBER-dated recessions.}
		\label{cumloss2added}
\end{figure}

\begin{figure}[H]
\centering
		\includegraphics[width=\textwidth]{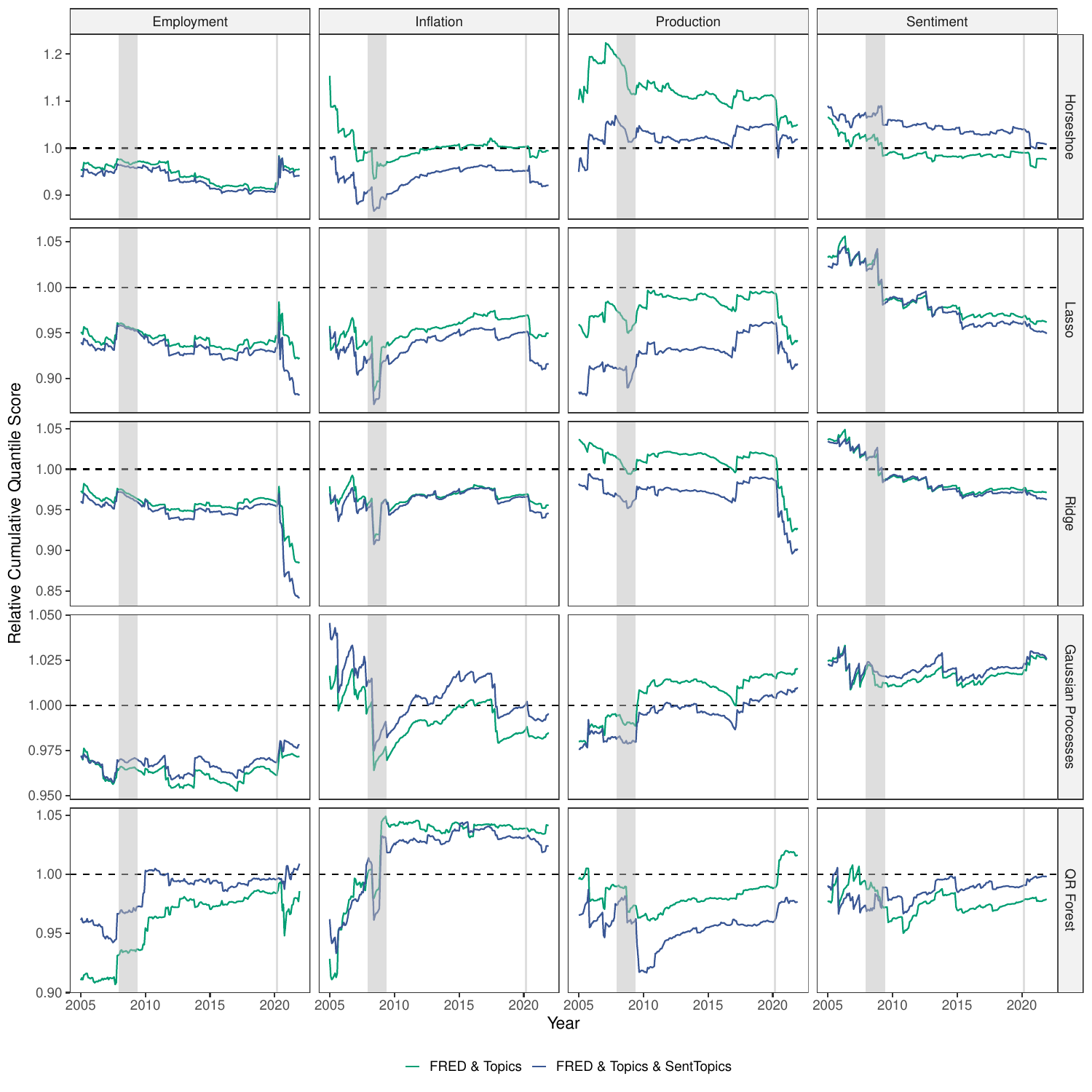} 
		\caption{\footnotesize One month ahead (h = 1) relative cumulative quantile score for the $90\%$-quantile ($\tau=0.9$) relative to the respective model and the FRED only benchmark.   The cumulative quantile scores of the respective benchmarks are standardized to 1.0. Scores below (above) 1.0 indicate more (less) precise forecasts compared to the benchmark until the given point in time. Gray-shaded areas indicate NBER-dated recessions.}
		\label{cumloss3added}
\end{figure}

\newpage

\section{Variable transformations}\label{AppC}
\begin{table}[h!]
		\centering
	%	\caption{Description of the Dataset} \label{tab: dataset}
		\scalebox{0.80}{
			\begin{tabular}{lllcc} 
				
				\hline
				ID & FRED Code & Description & Transformation Codes & Financial\\ 
				\hline
				1 & RPI & Real Personal Income & 5 & \\
				2 & W875RX1 & Real personal income ex transfer receipts & 5 & \\
				4 & CMRMTSPLx & Real Manu. and Trade Industries Sales & 5 & \\
				5 & RETAILx & Retail and Food Services Sales & 5 & \\
				6 & INDPRO & IP Index & 5 & \\
				7 & IPFPNSS & IP: Final Products and Nonindustrial Supplies & 5 & \\
				8 & IPFINAL & IP: Final Products (Market Group) & 5 & \\
				9 & IPCONGD & IP: Consumer Goods & 5 & \\
				13 & IPMAT & IP: Materials & 5 & \\
				16 & IPMANSICS & IP: Manufacturing (SIC) & 5 & \\
				20 & CUMFNS & Capacity Utilization: Manufacturing & 2 & \\
				23 & CLF16OV & Civilian Labor Force & 5 & \\
				24 & CE16OV & Civilian Employment & 5 & \\
				25 & UNRATE & Civilian Unemployment Rate & 2 & \\
				26 & UEMPMEAN & Average Duration of Unemployment (Weeks) & 2 & \\
				27 & UEMPLT5 & Civilians Unemployed - Less Than 5 Weeks & 5 & \\
				28 & UEMP5TO14 & Civilians Unemployed for 5-14 Weeks & 5 & \\
				29 & UEMP15OV & Civilians Unemployed - 15 Weeks \& Over & 5 & \\
				30 & UEMP15T26 & Civilians Unemployed for 15-26 Weeks & 5 & \\
				31 & UEMP27OV & Civilians Unemployed for 27 Weeks and Over & 5 & \\
				32 & CLAIMSx & Initial Claims & 5 & \\
				33 & PAYEMS & All Employees: Total nonfarm & 5 & \\
				34 & USGOOD & All Employees: Goods-Producing Industries & 5 & \\
				35 & CES1021000001 & All Employees: Mining and Logging: Mining & 5 & \\
				36 & USCONS & All Employees: Construction & 5 & \\
				37 & MANEMP & All Employees: Manufacturing & 5 & \\
				38 & DMANEMP & All Employees: Durable goods & 5 & \\
				39 & NDMANEMP & All Employees: Nondurable goods & 5 & \\
				40 & SRVPRD & All Employees: Service-Providing Industries & 5 & \\
				42 & USWTRADE & All Employees: Wholesale Trade & 5 & \\
				43 & USTRADE & All Employees: Retail Trade & 5 & \\
				44 & USFIRE & All Employees: Financial Activities & 5 & \\
				45 & USGOVT & All Employees: Government & 5 & \\
				46 & CES0600000007 & Avg Weekly Hours : Goods-Producing & 1 & \\
				47 & AWOTMAN & Avg Weekly Overtime Hours : Manufacturing & 2 & \\
				48 & AWHMAN & Avg Weekly Hours : Manufacturing & 1 & \\
				50 & HOUST & Housing Starts: Total New Privately Owned & 4 & \\
				51 & HOUSTNE & Housing Starts, Northeast & 4 & \\
				52 & HOUSTMW & Housing Starts, Midwest & 4 & \\
				53 & HOUSTS & Housing Starts, South & 4 & \\
				54 & HOUSTW & Housing Starts, West & 4 & \\
				55 & PERMIT & New Private Housing Permits (SAAR) & 4 & \\
				56 & PERMITNE & New Private Housing Permits, Northeast (SAAR) & 4 & \\
				57 & PERMITMW & New Private Housing Permits, Midwest (SAAR) & 4 & \\
				58 & PERMITS & New Private Housing Permits, South (SAAR) & 4 & \\
				59 & PERMITW & New Private Housing Permits, West (SAAR) & 4 & \\
				65 & AMDMNOx & New Orders for Durable Goods & 5 & \\
				66 & ANDENOx & New Orders for Nondefense Capital Goods & 5 & \\
				67 & AMDMUOx & Un lled Orders for Durable Goods & 5 & \\
				68 & BUSINVx & Total Business Inventories & 5 & \\
				69 & ISRATIOx & Total Business: Inventories to Sales Ratio & 2 & \\
				70 & M1SL & M1 Money Stock & 6 & \\
				\hline
			\end{tabular}
		}
	\end{table}

	\begin{table}[h!]
		\centering
		\scalebox{0.8}{
			\begin{tabular}{lllcc} 
				
				\hline
				ID & FRED Code & Description & Transformation Codes & Financial\\ 
				\hline
				71 & M2SL & M2 Money Stock & 6 & \\
				72 & M2REAL & Real M2 Money Stock & 5 & \\
				74 & TOTRESNS & Total Reserves of Depository Institutions & 6 & \\
				75 & NONBORRES & Reserves Of Depository Institutions & 7 & \\
				76 & BUSLOANS & Commercial and Industrial Loans & 6 & \\
				77 & REALLN & Real Estate Loans at All Commercial Banks & 6 & \\
				78 & NONREVSL & Total Nonrevolving Credit & 6 & \\
				79 & CONSPI & Nonrevolving consumer credit to Personal Income & 2 & \\
				80 & S\&P 500 & S\&P s Common Stock Price Index: Composite & 5 & X\\
				81 & S\&P: indust & S\&P s Common Stock Price Index: Industrials & 5 & X \\
				82 & S\&P div yield & S\&P s Composite Common Stock: Dividend Yield & 2 & \\
				83 & S\&P PE ratio & S\&P s Composite Common Stock: Price-Earnings Ratio & 5 & \\
				84 & FEDFUNDS & Effective Federal Funds Rate & 2 & X \\
				86 & TB3MS & 3-Month Treasury Bill: & 2 & X \\
				87 & TB6MS & 6-Month Treasury Bill: & 2 & X \\
				88 & GS1 & 1-Year Treasury Rate & 2 & X \\
				89 & GS5 & 5-Year Treasury Rate & 2 & X \\
				90 & GS10 & 10-Year Treasury Rate & 2 & X \\
				91 & AAA & Moody s Seasoned Aaa Corporate Bond Yield & 2 & X \\
				92 & BAA & Moody s Seasoned Baa Corporate Bond Yield & 2 & X \\
				94 & TB3SMFFM & 3-Month Treasury C Minus FEDFUNDS & 1 & X \\
				95 & TB6SMFFM & 6-Month Treasury C Minus FEDFUNDS & 1 & X \\
				96 & T1YFFM & 1-Year Treasury C Minus FEDFUNDS & 1 & X \\
				97 & T5YFFM & 5-Year Treasury C Minus FEDFUNDS & 1 & X \\
				98 & T10YFFM & 10-Year Treasury C Minus FEDFUNDS & 1 & X \\
				99 & AAAFFM & Moody s Aaa Corporate Bond Minus FEDFUNDS & 1 & X \\
				100 & BAAFFM & Moody s Baa Corporate Bond Minus FEDFUNDS & 1 & X \\
				102 & EXSZUSx & Switzerland / U.S. Foreign Exchange Rate & 5 & X \\
				103 & EXJPUSx & Japan / U.S. Foreign Exchange Rate & 5 & X \\
				104 & EXUSUKx & U.S. / U.K. Foreign Exchange Rate & 5 & X \\
				105 & EXCAUSx & Canada / U.S. Foreign Exchange Rate & 5 & X \\
				110 & OILPRICEx & Crude Oil, spliced WTI and Cushing & 6 & \\
				111 & PPICMM & PPI: Metals and metal products: & 6 & \\
				113 & CPIAUCSL & CPI : All Items & 6 & \\
				114 & CPIAPPSL & CPI : Apparel & 6 & \\	
				115 & CPITRNSL & CPI : Transportation & 6 & \\
				116 & CPIMEDSL & CPI : Medical Care & 6 & \\
				117 & CUSR0000SAC & CPI : Commodities & 6 & \\
				118 & CUSR0000SAD & CPI : Durables & 6 & \\
				119 & CUSR0000SAS & CPI : Services & 6 & \\
				120 & CPIULFSL & CPI : All Items Less Food & 6 & \\
				121 & CUSR0000SA0L2 & CPI : All items less shelter & 6 & \\
				122 & CUSR0000SA0L5 & CPI : All items less medical care & 6 & \\
				123 & PCEPI & Personal Cons. Expend.: Chain Index & 6 & \\
				127 & CES0600000008 & Avg Hourly Earnings : Goods-Producing & 6 & \\
				128 & CES2000000008 & Avg Hourly Earnings : Construction & 6 & \\
				129 & CES3000000008 & Avg Hourly Earnings : Manufacturing & 6 & \\
				130 & UMCSENTx & Consumer Sentiment Index & 2 & \\
				\hline
			\end{tabular}
		}
	\smallskip
	\begin{minipage}{\linewidth}\footnotesize
	 \textbf{Notes}: This table provides an overview of the \cite{McCracken2016} \textit{FRED-MD} data set. The transformation codes are applied to each time series $Y_j$ and described in : (1) no transformation; (2) $\Delta y_{jt}$; (3) $\Delta^2 y_{jt}$; (4) $\log (y_{jt})$; (5) $\Delta \log (y_{jt})$; (6) $\Delta^2 \log (y_{jt})$; (7) $\Delta (y_{jt}/y_{jt-1} - 1)$. Financial variables are indicated by X.
	\end{minipage}

	\end{table}

\end{document}